\documentclass[12pt]{article}
\usepackage{psfrag,amsmath,euscript,amssymb,graphicx}

\allowdisplaybreaks

\newcommand{\be}{\begin{equation}}
\newcommand{\ee}{\end{equation}}
\newcommand{\bea}{\begin{eqnarray}}
\newcommand{\eea}{\end{eqnarray}}
\newcommand{\nl}{\nonumber\\}
\newcommand{\order}{{\cal O}}

\newcommand{\sla}[1]{\rlap{\hspace{0.02cm}/}{#1}}
\newcommand{\ov}[1]{ \overleftarrow{#1} }
\def\nb{\bar{n}}
\def\calAslash{\rlap{\hspace{0.08cm}/}{{\EuScript A}}}

\def\bm#1{\mbox{\boldmath$#1$\unboldmath}}

\def\A{{\EuScript A}}

\def\H{{\EuScript H}}
\def\Q{{\EuScript Q}}

\def\X{{\EuScript X}}
\def\J{{\EuScript J}}

\begin{document}

\begin{titlepage}

\begin{flushright}
SLAC-PUB-10680\\
{\tt hep-ph/0408344}\\[0.2cm]
August 30, 2004
\end{flushright}

\vspace{0.7cm}
\begin{center}
\Large\bf
Loop Corrections to Heavy-to-Light Form Factors 
and Evanescent Operators in SCET
\end{center}

\vspace{0.8cm}
\begin{center}
{\sc  Thomas Becher and Richard J. Hill }\\
\vspace{0.7cm}
{\sl Stanford Linear Accelerator Center, Stanford University\\
Stanford, CA 94309, U.S.A.} \\
\end{center}

\vspace{1.0cm}
\begin{abstract}
\vspace{0.2cm}\noindent One-loop matching corrections are calculated
for Soft-Collinear Effective Theory (SCET) operators relevant to the
analysis of heavy-to-light meson form factors at large recoil.  The
numerical impact of radiative corrections on form factor predictions
is assessed. Evanescent operators in the effective theory are studied
and it is shown that even in problems of the Sudakov type, these
operators can be renormalized to have vanishing matrix elements.
\end{abstract}
\vfil

\end{titlepage}

\section{Introduction}

The methods that were used to establish factorization theorems for
high-energy QCD processes can also be applied to study decays of
$B$-mesons into light hadrons. In this case the hard scale is set by
the mass of the heavy $b$-quark and factorization theorems for
inclusive \cite{Neubert:1993ch, Bigi:1993ex, Korchemsky:1994jb} as
well as exclusive \cite{Beneke:1999br} $B$-decays to light hadrons
arise in the heavy-quark limit.  Soft-Collinear Effective Theory
(SCET) is the effective field theory arising in this limit. It describes the
heavy-quark expansion for these decays, and permits the study of their factorization
and renormalization properties
\cite{Bauer:2000yr,Bauer:2001yt,Chay:2002vy,Beneke:2002ph,Hill:2002vw}.

Semileptonic decays, such as $B\rightarrow \pi \ell \nu$, 
are the simplest exclusive heavy-to-light meson processes. 
These decays are described by weak-interaction form factors, 
which at large recoil energies of the light meson, $E\sim m_B/2 \gg \Lambda_{\rm QCD}$, 
take the form
\cite{Beneke:2000wa},
\begin{equation}\label{eq:soft_plus_hard}
  F_i^{B\to M}(E) = C_i(E)\,\zeta_M(E) +
  \int_0^\infty\!{d\omega\over\omega}\,\phi_B(\omega)
  \int_0^1\!du\,f_M\,\phi_M(u)\,T_i(E,\omega,u) \,,
\end{equation}
up to corrections suppressed by $\Lambda_{\rm QCD}/m_b$. 
Here $M$ denotes the final state (pseudoscalar or vector) meson.  
The process-dependent Wilson coefficient functions $C_i(E)$ and $T_i(E,\omega,u)$ 
are calculable in perturbation theory and are the subject of the present
paper.  The wave functions, or more precisely, light cone distribution amplitudes (LCDAs)
 $\phi_B(\omega)$ and $\phi_M(u)$, and
the functions  $\zeta_M(E)$, are process-independent nonperturbative hadronic parameters. 
The factorization theorem (\ref{eq:soft_plus_hard}) has recently been established 
in the context of SCET~\cite{Bauer:2002aj,Beneke:2003pa,Lange:2003pk}. 

At large recoil, the form factors $F_i^{B\to M}$ are in fact subleading quantities, 
requiring power-suppressed interactions to mediate the $B\to M$ transition.  
Since the light degrees of freedom inside
the $B$-meson carry soft momenta, $p^\mu \sim \Lambda_{\rm QCD}$, 
whereas the degrees of freedom inside the final-state meson typically share 
the large energy of the meson, the transition 
can only occur if:  (i), the partons are in an atypical ``endpoint'' configuration, 
allowing the soft $B$-meson constituents to be absorbed into the energetic light meson; 
or (ii), a large momentum is exchanged between the active quark and the light spectator degrees of freedom.    
Possibility (i), the so-called soft-overlap mechanism, 
is described by the non-factorizable part $\zeta_M(E)$ of the form factors,
and in the effective theory is given by the matrix element of the leading-order
heavy-to-light quark current.  This current involves only the two-component spinor
fields of SCET and Heavy-Quark Effective Theory (HQET)~\cite{Neubert:1993mb}, 
and mediates the transition of the  heavy quark inside the $B$-meson
into an energetic light quark component of the light meson.  
The simple spin structure of the two-component fields results in ``large-energy'' spin symmetries, 
such that a single function $\zeta_M(E)$ 
describes the soft-overlap contribution to all form factors involving the same final-state meson 
$M$~\cite{Charles:1998dr}.    
Possibility (ii), the hard-scattering mechanism, 
results from subleading effective theory current operators involving an additional gluon field transferring
large momentum to the spectator.  Their matrix elements
factorize and correspond to the second term in equation
(\ref{eq:soft_plus_hard}).

\begin{figure}[htb]
\begin{center}
\includegraphics[width=0.9\textwidth]{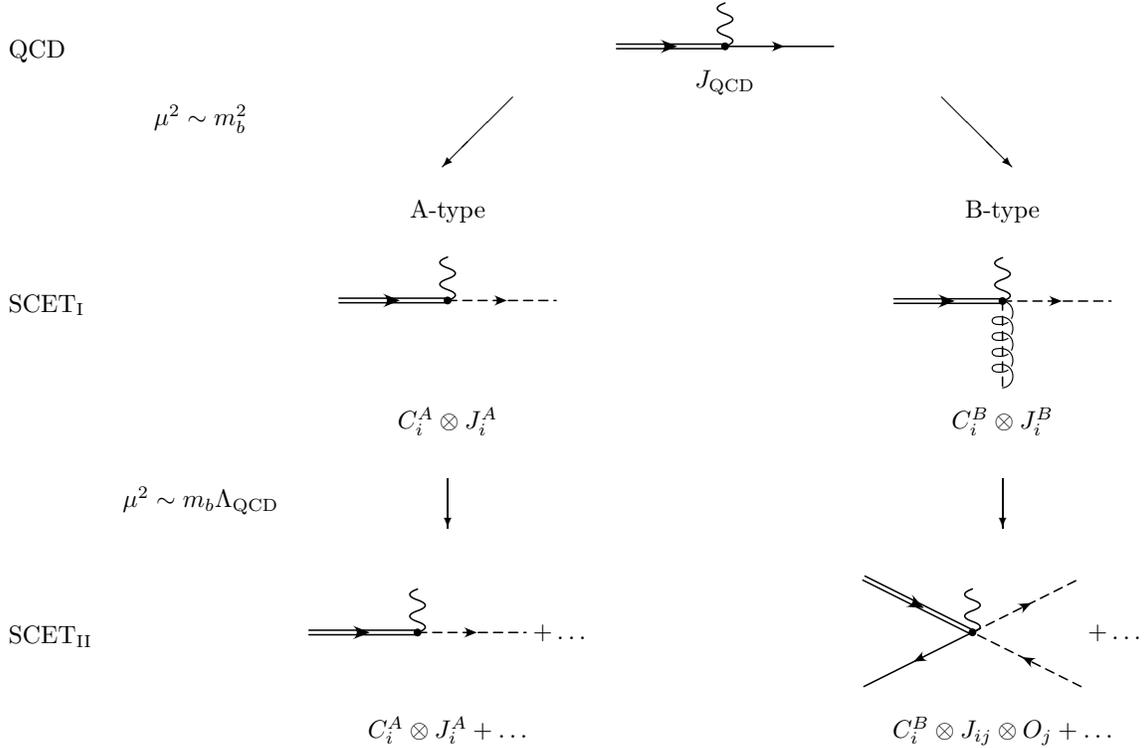}
\end{center}
\vspace{-0.5cm}
\caption{Two step matching from QCD to SCET$_{\rm I}$ to SCET$_{\rm
II}$. The double line denotes the heavy quark; dashed lines
denote hard-collinear fields in SCET$_{\rm I}$ and collinear fields in
SCET$_{\rm II}$. The symbol ``$\otimes$'' indicates the
convolution of the Wilson coefficient functions with the non-local
operators. }
\label{fig:matching}
\end{figure}

The form factors receive contributions associated with different
momentum scales: the hard scale $\mu^2\sim m_b^2$, an intermediate
scale $\mu^2\sim m_b\Lambda_{\rm QCD}$ and the soft scale
$\mu^2\sim\Lambda_{\rm QCD}^2$. In order to perform a
renormalization-group (RG) resummation of large perturbative
logarithms involving ratios of these scales, a two-step matching is
required, first from QCD onto an intermediate theory, denoted
SCET$_{\rm I}$, and then from SCET$_{\rm I}$ onto the final low-energy
theory, denoted SCET$_{\rm II}$; see Figure~\ref{fig:matching}.  For
the leading-power currents defining the soft-overlap contributions,
the matching from QCD onto SCET$_{\rm I}$ has been performed through
one-loop order in %
\cite{Bauer:2000yr}, 
and the subsequent matching
onto SCET$_{\rm II}$ is given in 
\cite{Lange:2003pk}.  
Resummation of large logarithms in the coefficient functions $C_i(E)$ results in a
universal RG factor%
~\cite{Bauer:2000yr,Lange:2003pk}, 
the same for all form factors. 
However, the numerical value of this universal factor is presently of
limited utility, since the functions $\zeta_M(E)$ are not known at any
fixed renormalization scale.  On the other hand, some predictions
exist for the hadronic wavefunctions appearing in the second term of
(\ref{eq:soft_plus_hard}), and we may study the numerical impact of
renormalization-group evolution down to hadronic scales where these
predictions are reliable.  In a recent paper in collaboration with Lee
and Neubert~%
\cite{Hill:2004if}, 
we resummed all leading single and
double logarithms in the hard scattering kernels $T_i(E,\omega,u)$.
At leading order in RG-improved perturbation theory, the resummed
result is obtained using tree-level matching.  However, since
$\alpha_s$ is sizeable at the relevant scales, we expect one-loop
matching corrections to also be important. In 
\cite{Hill:2004if} 
the size of these matching corrections was investigated using the scalar
current as an example, and they were found to be comparable to the
resummation effects.  In the present paper, we compute the complete
one-loop matching results for the hard-scattering term, for both
matching steps.

The outline of the paper is as follows. 
In Section~\ref{sec:scetI} we give the matching coefficients for 
QCD vector and tensor currents onto the subleading SCET$_{\rm I}$ operators relevant to the
hard-scattering form factor contributions.
The coefficients for the scalar current have been calculated previously~%
\cite{Hill:2004if,Beneke:2004rc}, 
and can in fact be derived from the vector current results~%
\cite{Hill:2004if}.     
The vector and tensor coefficients have also 
been calculated by Beneke, Kiyo and Yang~%
\cite{Beneke:2004rc} 
and we find full agreement with their results; in 
Section~\ref{sec:scetI} we list only those coefficients necessary for 
$B\to M$ form factors, in the ``primed'' operator basis of 
\cite{Hill:2004if}.  
The Wilson coefficients, also called jet functions, that arise in the second matching step, from
SCET$_{\rm I}$ onto SCET$_{\rm II}$, are
considered in Section~\ref{sec:scetII}.  We have already stated the results
relevant to the form factors in 
\cite{Hill:2004if}.
Here we present the general expressions from which those results were
derived, and provide details of the calculation.

For dimension $d=4-2\epsilon$ in dimensional regularization, so-called
evanescent operators~%
\cite{Buras:1989xd,Dugan:1990df,Herrlich:1994kh}
arise in the matching of QCD onto SCET.  
The matrix elements of these operators vanish at tree-level for $d=4$, but
in a generic renormalization scheme they are nonzero at higher order.
To avoid introducing new hadronic functions at each order in
perturbation theory to parameterize the matrix elements of the
evanescent operators, it is important to show that a renormalization
scheme exists for which these operators vanish exactly in four
dimensions.  The existence of such a scheme is related to the
technical observation that in loop Feynman diagrams involving an
evanescent operator, a contribution proportional to a physical
operator contains a factor $\epsilon$ from Dirac algebra, which
cancels a $1/\epsilon$ divergence arising from integration over loop
momenta; the resulting finite contribution is local, and so may be
subtracted by a suitable counterterm.  It can also be shown that in
this scheme no mixing occurs of the physical into the evanescent
operators, and thus the matching coefficients onto renormalized evanescent
operators are irrelevant to the calculation of physical matrix
elements.  In processes involving Sudakov double logarithms, the
existence of $1/\epsilon^2$ divergences in loop integrations poses a
potential obstacle to these standard arguments.  In
Section~\ref{sec:evanescent} we show that despite this complication, 
renormalization schemes exist 
in which the matrix elements of
renormalized evanescent operators vanish.
This section also discusses a related issue, 
the choice of evanescent operators.  
Different choices correspond to different renormalization schemes for
the physical operators. 
For the four-quark SCET$_{\rm II}$ operators describing the 
hard-scattering form factor contributions, 
we isolate the particular operator basis which corresponds to the 
$\overline{\rm MS}$ scheme after Fierz transformation.  Upon taking 
heavy-light meson matrix elements, this ensures that the LCDAs appearing
in (\ref{eq:soft_plus_hard}) are renormalized in the $\overline{\rm MS}$ scheme.

The phenomenological importance of one-loop matching corrections to
heavy-to-light form factors is investigated in Section~\ref{sec:app}.
Hard-scale corrections, corresponding to QCD $\rightarrow$ SCET$_{\rm I}$ 
matching, are found to be of order $10-20\%$.  These corrections
determine the size of violations to the large-energy spin-symmetry
relations obeyed by the soft-overlap term, and also the size of
non-universal corrections to the hard-scattering term.  We notice that
in the soft-overlap case, the radiative corrections for different form
factors are remarkably similar, and therefore have little effect on
the symmetry relations.  The numerical impact of one-loop
contributions to the jet functions for SCET$_{\rm I}$ $\rightarrow$
SCET$_{\rm II}$ matching is more difficult to estimate, owing to the
lack of precise knowledge on the $B$-meson wavefunction.  However,
when hard-scale QCD $\rightarrow$ SCET$_{\rm I}$ matching corrections
are ignored, the hard-scattering contributions are described by the
universal functions $H_M(E)$ introduced in 
\cite{Hill:2004if};
radiative corrections to the jet functions are therefore not relevant
to the conclusions which can be drawn based solely on this
universality.  The radiative corrections {\it are} necessary in order
to relate the functions $H_M(E)$ to models for the $B$-meson
wavefunction, and we investigate the size of the corrections for two
such models, finding that they increase $H_M(E)$ by $\sim 20-30\%$.
Section~\ref{sec:conclusion} provides a discussion and our
conclusions.

\section{One-loop matching from QCD onto SCET$_{\bm{\rm I}}$\label{sec:scetI}}

In this section, we deal with 
SCET$_{\rm I}$, which describes the interaction of energetic light particles 
of virtuality $p^2 \sim m_b \Lambda_{\rm QCD}$ with a static heavy quark. 
The soft sector of SCET$_{\rm I}$ is described by HQET, 
with a heavy quark of velocity $v$.       
Different components of hard-collinear momenta and fields scale with 
different powers of the expansion parameter $\lambda^{1/2}=(\Lambda_{\rm QCD}/m_b)^{1/2}$. 
The large momentum components are isolated to obtain a definite power counting 
by working with light-like reference vectors, $n^\mu$ in the direction of the jet of outgoing
collinear particles, and $\nb^\mu$ a complementary vector satisfying $n\cdot\nb=2$:      
\begin{equation}
   p^\mu = n\cdot p\,{\nb^\mu\over 2} + \nb\cdot p\,{n^\mu\over 2} +
    p_{\perp}^\mu \,.  
\end{equation}
The components $(n\cdot p,\nb\cdot p, p_\perp)$ of a hard-collinear momentum are defined to scale as
$(\lambda,1,\lambda^{1/2})$, while for a soft momentum the components scale as 
$(\lambda,\lambda,\lambda)$.  
We choose $\nb^\mu$ such that $v_\perp=0$, implying $\nb^\mu=(2v^\mu-n^\mu/n\cdot v)/n\cdot v$.    
The canonical choices are 
$v^\mu=(1,0,0,0)$, $n^\mu=(1,0,0,1)$ and $\nb^\mu=(1,0,0,-1)$, corresponding to 
the energetic jet in the $z$-direction as viewed from the rest-frame 
of the heavy meson.   
Throughout this paper we use the notation and conventions of \cite{Hill:2004if}, to which we refer the
reader for details and references.

We now consider the effective theory representation of the scalar, vector
and tensor flavor-changing currents, 
\begin{equation}
 S = \bar{q}\,\, b\,, \qquad V^\mu = \bar q\,\gamma^\mu \,b \,, \qquad
  T^{\mu\nu} = (-i)\,\bar q\,\sigma^{\mu\nu}\,b = \bar
  q\,\gamma^{[\mu}\gamma^{\nu]}\,b \,.
\end{equation}
We use square brackets around indices to denote antisymmetrization. 
The QCD scalar and tensor currents require renormalization. 
We work in dimensional regularization for $d=4-2\epsilon$ dimensions, 
and define the QCD as well as the effective theory operators in the modified
minimal subtraction ($\overline{\rm MS}$) scheme.%
\footnote{
In Section~\ref{sec:evanescent}, we consider the modifications to this scheme necessary 
to ensure the vanishing of renormalized evanescent operators. 
}
We use the naive
dimensional regularization (NDR) scheme for $\gamma_5$ so that the
pseudoscalar and pseudovector currents renormalize identically to the 
scalar and vector currents, and our
results for the matching are independent of the chirality of the light
quark $q$ in the current operator. 
The construction of
the effective theory currents is discussed in 
\cite{Hill:2004if}.
For the case of the scalar current, which we take at position $x=0$, 
only two effective theory operators are needed through the first subleading order:
\begin{align}\label{eq:scalar_currents}
   J^A_S(s,x=0) &= \bar\X_{hc}(s\nb) \bigg( 1
    - {i\ov{\sla{\partial}}_{\!\perp}\over i\nb\cdot\ov{\partial}}\,
    {\sla{\nb}\over 2}  \bigg) h(0) \,, \nl
   J^B_S(s,r,x=0) &= 
    \bar\X_{hc}(s\nb)\,\calAslash_{hc\perp}(r\nb)\,h(0) \,.
\end{align}
The field $h$ is the heavy-quark field, $\X_{hc}=W_{hc}^\dagger\xi_{hc}$ is the
hard-collinear quark field multiplied by a hard-collinear Wilson line
$W_{hc}$ in the $\nb$-direction, and 
$\A_{hc\perp}^\mu = W_{hc}^\dagger\,[(i\partial_{\perp}^\mu+A_{hc\perp}^\mu)
W_{hc}]$, where $A_{hc}$ is the hard-collinear gluon field. 
Note that the above operators are
non-local: the hard-collinear fields live at different points on the
light-ray in the $\nb$-direction. This is a consequence of the fact that
$\nb\cdot \partial$ derivatives on hard-collinear fields count as
quantities of order one, so that operators involving arbitrary functions
of such derivatives may appear at a given order in the power counting. 
The Wilson coefficients of these operators
are then functions of the position arguments and the representation of the
scalar current at $x=0$ in the effective theory reads
\begin{eqnarray}\label{Jexpansion} 
  \bar q\,b &\to&
    \int ds\,\tilde C_S^A(s)\,J_S^A(s) + \frac{1}{2E} \int dr\,ds\,\tilde
    C_S^B(s,r)\,J_S^B(s,r) + \dots \nl &=& C^A_S(E)\,J^A_S(0) +
\frac{1}{2E} \int du\,C^B_S(E,u)\,J^B_S(u) + \dots \,, 
\end{eqnarray}
where the factor $1/2E$ has been inserted into the second term for convenience. 
We use a tilde to denote the coefficients in position space. The
momentum space coefficient functions in the second line are defined as
\begin{eqnarray} 
  C^A_S(E) &=& \int ds\,e^{is\nb\cdot P}\,\tilde C^A_S(s) \,, \nl
  C^B_S(E,u) &=& \int dr\,ds\,e^{i(us+\bar u r)\nb\cdot P}\, \tilde
    C^B_S(s,r) \,.  
\end{eqnarray}
Here $P=P_{\rm out}-P_{\rm in}$ is the total
hard-collinear momentum of external states, and $E\equiv(n\cdot v)(\bar n\cdot P)/2$. 
The variable $u\in [0,1]$ denotes the fraction of the energy $E$ carried by the 
quark field $\bar{\X}_{hc}$ and $\bar{u}=1-u$ the fraction carried by $\A_{hc\perp}^\mu$.

The fact that the subleading term in $J^A_S$ involving a
perpendicular partial derivative has the same Wilson coefficient as the
leading order term is a consequence of reparametrization invariance,
the fact that QCD is independent of the reference vectors $n_\mu$ and
$v_\mu$ used in the construction of the effective theory~%
\cite{Chay:2002vy,Manohar:2002fd,Hill:2004if}. 
Due to
the presence of the reference vectors $n_\mu$ and $v_\mu$, the vector
and tensor currents are represented by several effective theory
operators, even at leading order. However the general structure of the
subleading currents is the same as in the scalar case: $A$-type
operators involve perpendicular partial derivatives and have the same Wilson
coefficient as the leading order operators;  the remaining, $B$-type
operators involve the insertion of an additional perpendicular gluon
field $\A_{hc\perp}^\mu$.  The complete basis of SCET$_{\rm I}$ current
operators can be found e.g.~in 
\cite{Hill:2004if}. 
In that reference two different sets of basis operators are used, $J^B_i$ and
$J^{B'}_{i}$. Here we use the $B$-type operators in the primed basis, which are
more convenient in that they renormalize multiplicatively and match
in a simple way onto SCET$_{\rm II}$. Suppressing the ``$hc$'' subscript on hard-collinear fields,
the vector current operators in this basis are
\begin{align}\label{eq:vectorprime}
  J_{V1}^{B'\mu} &= \bar\X(s\nb)\,\calAslash_\perp(r\nb)\,\gamma_\perp^\mu\,h(0) \,, &
  J_{V2}^{B'\mu} &= \bar\X(s\nb)\,\calAslash_\perp(r\nb)\,v^\mu\,h(0) \,, \nl
  J_{V3}^{B'\mu} &= \bar\X(s\nb)\,\calAslash_\perp(r\nb)\,\frac{n^\mu}{n\cdot v}\,h(0) \,, &
  J_{V4}^{B'\mu} &= \bar\X(s\nb)\,\gamma_\perp^\mu\,\calAslash_\perp(r\nb)\,h(0) \,,
\end{align}
and the tensor current is represented by
\begin{align}\label{eq:tensorprime}
  J_{T1}^{B'\mu\nu} &= \bar\X(s\nb)\,\calAslash_\perp(r\nb)\,\gamma_\perp^{[\mu}\gamma_\perp^{\nu]}\,h(0) \,, &
  J_{T2}^{B'\mu\nu} &= \bar\X(s\nb)\,\calAslash_\perp(r\nb)\,v^{[\mu}\gamma_\perp^{\nu]}\,h(0)\,, \nl
  J_{T3}^{B'\mu\nu} &= \bar\X(s\nb)\,\calAslash_\perp(r\nb)\,{n^{[\mu}\gamma_\perp^{\nu]}\over n\cdot v}\,h(0) \,, &
  J_{T4}^{B'\mu\nu} &= \bar\X(s\nb)\,\calAslash_\perp(r\nb)\,{n^{[\mu}v^{\nu]}\over n\cdot v}\,h(0) \,,\nl
  J_{T5}^{B'\mu\nu} &= \bar\X(s\nb)\,\A_{\perp\alpha}(r\nb)\,\gamma_\perp^{[\alpha}\gamma_\perp^\mu\gamma_\perp^{\nu]}\,h(0) \,, &
  J_{T6}^{B'\mu\nu} &= \bar\X(s\nb)\,v^{[\mu}\gamma_\perp^{\nu]}\,\calAslash_\perp(r\nb)\,h(0) \,, \nl
  J_{T7}^{B'\mu\nu} &= \bar\X(s\nb)\,\frac{n^{[\mu}\gamma_\perp^{\nu]}}{n\cdot v}\,\calAslash_\perp(r\nb)\,h(0) \,. &
\end{align}

\begin{figure}
\begin{center}
\includegraphics[width=0.75\textwidth]{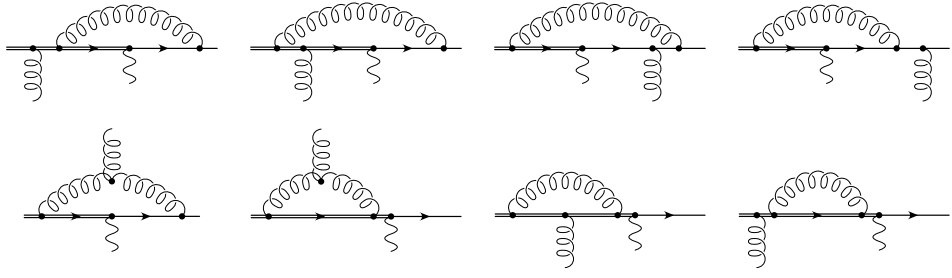}
\end{center}
\vspace{-0.5cm}
\caption{One-loop QCD diagrams contributing to the matching calculation for
the subleading scalar current. The double line denotes the heavy quark.}
\label{fig:full}
\end{figure}

One-loop matching corrections to the leading order, A-type effective theory
currents were evaluated in 
\cite{Bauer:2000yr} 
and the results were
confirmed by the findings of 
\cite{Beneke:2004rc}. 
Here, we focus on the Wilson
coefficients of the $B$-type operators which can be obtained by evaluating
the QCD diagrams shown in Figure~\ref{fig:full}. By performing the
matching on-shell, one can avoid the calculation of effective theory
loop diagrams, as discussed in 
\cite{Hill:2004if},
where we presented the one-loop matching for the scalar current. 
Only six of the above currents match
onto SCET$_{\rm II}$ four-quark operators relevant for the
heavy-to-light form factors and we now give the
results for those six Wilson coefficients. 
Introducing $x=2E/m_b$, we find for the case of the vector current, 
\begin{eqnarray}\label{eq:vectorB} 
  C^{B'}_{V2} &=& -2 + {C_F\alpha_s\over 4\pi}\Bigg\{
    4\ln^2{2E\over\mu} - 2\ln{2E\over\mu} + 4{\rm Li}_2(1-x) + {\pi^2\over
    6} -{2\over u}\left[ {\ln{x}\over(1-x)^2}-
    {\ln{x\bar{u}}\over(1-x\bar{u})^2} \right.\nl 
  &&\qquad \left.  +
    {\ln{x}\over 1-x} - {\ln{x\bar{u}}\over 1-x\bar{u}} -2\ln{\bar{u}}
    \right]
    + {6\ln{x}\over 1-x}  
    -{2x\over (1-x)(1-x\bar{u})} 
    + 2 \Bigg\} \nl
  && +\left(C_F-{C_A\over 2}\right){\alpha_s\over 4\pi}\Bigg\{ 
    {4\over\bar{u}}\left[ 2\left(\ln{2E\over\mu}-1\right)\ln{u} +\ln^2{u} - {\rm Li}_2(1-x) + {\rm Li}_2(1-xu)\right]
    \nl
  && \qquad
    + {4\over u}\left[ {\ln{x}\over 1-x} - {\ln{x\bar{u}}\over 1-x\bar{u}}  - \ln\bar{u}\right]
    \Bigg\} \,,\nl 
  C^{B'}_{V3} &=& 1-x + {C_F\alpha_s\over 4\pi}\Bigg\{ 
    (1-x)\left[ -2\ln^2{2E\over\mu} + \ln{2E\over\mu} - 2{\rm Li}_2(1-x) - {\pi^2\over 12}\right]  \nl
  && \qquad 
    - \left[{1\over(1-x\bar{u})^2} +3 -2x  \right]{\ln{\bar{u}}\over u}  
    + \left[ {x\over (1-x\bar{u})^2} + {x\over (1-x)(1-x\bar{u}) } -3 \right]{\ln{x}\over 1-x} \nl
  && \qquad
    + {x\over (1-x)(1-x\bar{u})} -2 + x \Bigg\}  \nl
  && + \left(C_F-{C_A\over 2}\right){\alpha_s\over 4\pi}\Bigg\{ {2\over xu\bar{u}}\left[ {\rm Li}_2(1-x) 
    -{\rm Li}_2(1-xu) - {\rm Li}_2(1-x\bar{u}) + {\pi^2\over 6} \right] 
    \nl
  && \qquad
    + {2(2-x)\ln{u}\over\bar{u}} + 2\left[ {1\over 1-x\bar{u}} + 2- x\right]{\ln{\bar{u}}\over u} 
    - {2x\ln{x}\over(1-x)(1-x\bar{u})} \nl
  && \qquad
    -{2(1-x)\over\bar{u}}\left[ \left(2\ln{2E\over\mu} - 1\right)\ln{u} +\ln^2{u} - {\rm Li}_2(1-x)+{\rm Li}_2(1-xu) \right] 
    \Bigg\} \,,\nl
  C^{B'}_{V4} &=& {C_F\alpha_s\over 4\pi}{x\over 1-x\bar{u}}\Bigg\{  
    -\left[ {1\over 1-x\bar{u}} + {1\over u}\right]\ln{\bar{u}}  
    -\left[ {1\over 1-x\bar{u}} - {x\over 1-x}\right]\ln{x}  - 1 \Bigg\} \nl 
  && + \left(C_F-{C_A\over 2}\right){\alpha_s\over 4\pi}\Bigg\{ 
    -{2(1-x)\ln{u}\over \bar{u}(1-xu)} - {2(1-x)\ln{\bar{u}}\over u(1-x\bar{u})} 
    +{2x(2-x)\ln{x}\over (1-xu)(1-x\bar{u})} \nl
  && \qquad 
    -{2\over xu\bar{u}}\left[ {\rm Li}_2(1-x)-{\rm Li}_2(1-xu)-{\rm Li}_2(1-x\bar{u}) + {\pi^2\over 6}\right]\Bigg\} \,. 
\end{eqnarray}
The three coefficients needed for the tensor case are
\begin{eqnarray} \label{eq:tensorB}
  C^{B'}_{T4} &=& 2 + {C_F\alpha_s\over 4\pi}\Bigg\{ 
    -4\ln^2{2E\over\mu} + 2\ln{2E\over\mu} - 4\ln{\mu_{\rm QCD}\over 2E} - 4{\rm Li}_2(1-x) -  {\pi^2\over 6} \nl
  && \qquad\qquad\qquad + 2\left[ {1-x\over (1-x\bar{u})^2} - {2+x\over 1-x\bar{u}} - 3  \right]{\ln{\bar{u}}\over u} 
    \nl
  && \qquad
    - 2\left[ {x\over (1-x\bar{u})^2} - {x(1+x)\over (1-x)(1-x\bar{u}) } + 5  \right]\ln{x} 
    - {2x\over 1-x\bar{u}} -4 
    \Bigg\} \nl
  && 
    + \left(C_F-{C_A\over 2}\right){\alpha_s\over 4\pi}\Bigg\{ 
    -{4\over\bar{u}}\left[ \left(2\ln{2E\over \mu} - 1 \right){\ln{u}} + \ln^2{u} -{\rm Li}_2(1-x)+{\rm Li}_2(1-xu) \right] 
    \nl
  && \qquad\qquad
    - 4\left[{1-x\over 1-x\bar{u}} - 2 \right]{\ln{\bar{u}}\over u} 
    + {4x\ln{x}\over 1-x\bar{u}} 
    \nl
  && \qquad 
    -{4\over xu\bar{u}}\left[ {\rm Li}_2(1-x) -{\rm Li}_2(1-xu) - {\rm Li}_2(1-x\bar{u}) + {\pi^2\over 6} \right]
    \Bigg\} \,, \nl
  C^{B'}_{T6} &=& 0\,, \\
  C^{B'}_{T7} &=& 2x + {C_F\alpha_s\over 4\pi} x \Bigg\{ 
    - 4\ln^2{2E\over\mu} + 2\ln{2E\over\mu} - 4\ln{\mu_{\rm QCD}\over 2E} -4{\rm Li}_2(1-x)  - {\pi^2\over 6}
    \nl
  && \quad
    +{4\bar{u}\over u}\left[ \left(2\ln{2E\over\mu} - 1\right)\ln\bar{u}  + \ln^2\bar{u} -{\rm Li}_2(1-x)+{\rm Li}_2(1-x\bar{u}) \right]  
    -\frac{4}{u}\ln\bar{u}- 2\ln{x}  - 2  \Bigg\}  \nl
  && + \left(C_F-{C_A\over 2}\right){\alpha_s\over 4\pi} x \Bigg\{ 
    -{4\bar{u}\over u}\left[ \left(2\ln{2E\over\mu} - 1\right)\ln\bar{u}  + \ln^2\bar{u} -{\rm Li}_2(1-x)+{\rm Li}_2(1-x\bar{u}) \right] \nl
  && \qquad -{4(2-u)\over\bar{u}}\left[ \left(2\ln{2E\over \mu} - 1 \right){\ln{u}} 
    + \ln^2{u} -{\rm Li}_2(1-x)+{\rm Li}_2(1-xu) \right]  \nl
  && \qquad -{4\over xu\bar{u}}\left[ {\rm Li}_2(1-x)-{\rm Li}_2(1-xu)\right] 
    + {4\over xu}\left[ {\rm Li}_2(1-x\bar{u}) -{\pi^2\over 6} \right] 
    + 4\ln{\bar{u}}- 4\ln{u}  - 4 \Bigg\} \,.\nonumber
\end{eqnarray}
We have checked that our results for all four vector currents, 
and seven tensor currents (including those not listed here)
are in complete agreement with previous results of Beneke, Kiyo and Yang~%
\cite{Beneke:2004rc}
once we translate to their operator basis, using the relations
\begin{align}
  C^{B'}_{V1} &= C^{(A1)1}_V + \frac12 C^{(A1)4}_V + x\,C^{(B1)1}_V + \frac{x}{2}
  C^{(B1)4}_V\,, & 
  C^{B'}_{V2} &= -2\,C^{(A1)3}_V + x\,C^{(B1)3}_V\,,\nl 
  C^{B'}_{V3} &= C^{(A1)2}_V - x C^{(B1)2}_V \,,&
  C^{B'}_{V4} &= \frac12 C^{(A1)4}_V + \frac{x}{2} C^{(B1)4}_V\,,
\end{align}
for the vector coefficients, 
and for the tensor coefficients: 
\begin{align}
  C^{B'}_{T1} &= -C^{(A1)4}_T - x C^{(B1)4}_T \,, &
  C^{B'}_{T2} &= -4\,C^{(A1)2}_T + C^{(A1)6}_T - 2x\,C^{(B1)2}_T 
    + C^{(B1)6}_T\,,\nl 
  C^{B'}_{T4} &= 2\,C^{(A1)3}_3 - 2x\,C^{(B1)3}_T\,, & 
  C^{B'}_{T3} &= 2\,C^{(A1)1}_T + C^{(A1)5}_T - 2x\,C^{(B1)1}_T +
    2x\,C^{(B1)5}_T\,,\nl 
  C^{B'}_{T6} &= C^{(A1)6}_T + x\,C^{(B1)6}_T \,, &
  C^{B'}_{T5} &= C^{(A1)4}_T - C^{(A1)7}_T + x\,C^{(B1)4}_T - x\,C^{(B1)7}_T
    \,,\nl
  C^{B'}_{T7} &= C^{(A1)5}_T + 2x\,C^{(B1)5}_T \,. &
\end{align}

\section{One-loop matching from SCET$_{\bm{\rm I}}$ onto SCET$_{\bm{\rm II}}$\label{sec:scetII}}

\begin{figure}[t]
\begin{center}
\includegraphics[width=0.9\textwidth]{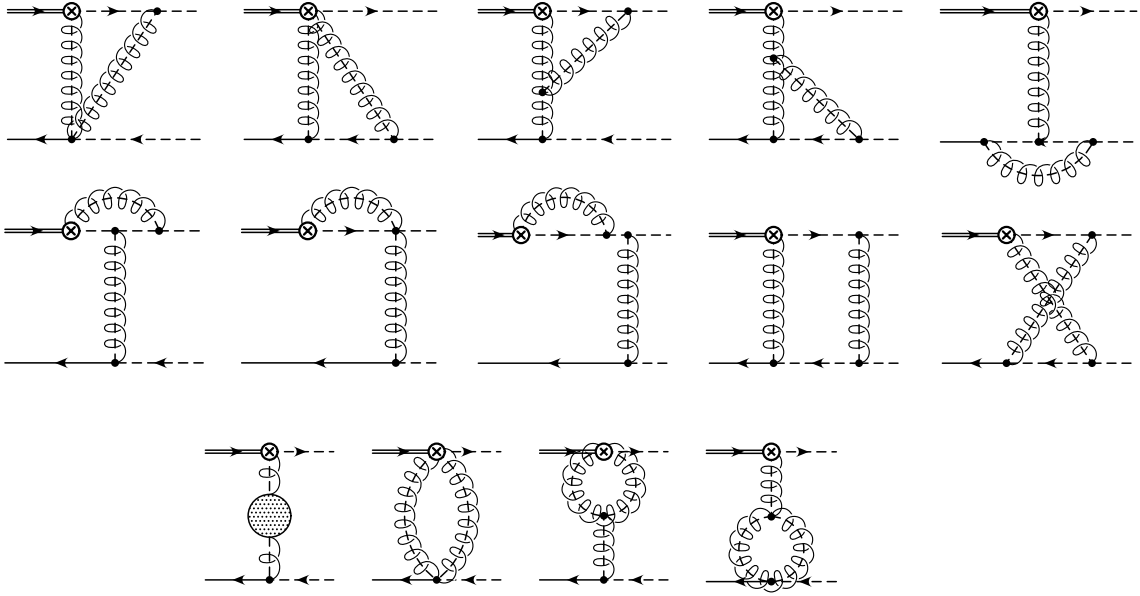}
\end{center}
\vspace{-0.5cm}
\caption{$\mbox{SCET}_{\rm I}$ graphs contributing to the matching of the 
currents $J_j^{B}$ (crossed circle) onto the $\mbox{SCET}_{\rm II}$ 
four-quark operators $O_i$. Full lines denote soft fields, dashed lines 
hard-collinear fields. Diagrams with soft gluons or scaleless loops are not 
shown.}
\label{fig:jet}
\end{figure}

We now map the $B$-type SCET$_{\rm I}$ currents onto SCET$_{\rm II}$
four-quark operators, as represented in the right-hand side of Figure~\ref{fig:matching}. 
We recall that SCET$_{\rm II}$ is the low-energy effective theory encompassing
collinear modes with momentum scaling $(n\cdot p, \nb\cdot p, p_\perp)\sim (\lambda^2,1,\lambda)$,
and soft modes with momentum scaling $(\lambda,\lambda,\lambda)$.  
In general, soft-collinear messenger 
modes, with momentum scaling $(\lambda^2,\lambda,\lambda^{3/2})$ are also required, even at leading order, and 
may communicate between the soft and collinear sectors~%
\cite{Becher:2003qh}.  
However, in the particular case of
the color-singlet four-quark operators under consideration,   
the sum over all such contributions can be shown to vanish;  
this is equivalent to the
statement that the soft and collinear sectors factorize for the corresponding matrix elements~%
\cite{Becher:2003kh,Lange:2003pk}.
The resulting factorizable contributions to the form factors take the form of the second term 
in (\ref{eq:soft_plus_hard}).   
The function $T_i(E,\omega,u)$ appearing in this formula is a product (in general, a convolution) of the 
SCET$_{\rm I}$ coefficients $C_i^{B}$ from the first matching step, considered in Section~\ref{sec:scetI}, 
and the Wilson coefficients occurring in the second matching step, called jet-functions. 
The fact that the heavy quark and the collinear quarks
are described by two component spinors
restricts the allowed form of possible SCET$_{\rm II}$ operators, and 
leads to relations among the jet-functions~%
\cite{Hill:2004if}. 
For the 
operators relevant to $B$-decay form factors, only two jet functions are 
relevant; these are denoted by ${\cal J}_{\parallel}$ and ${\cal J}_\perp$, and describe decays into
pseudoscalar or longitudinally-polarized vector mesons, and into perpendicularly-polarized 
vector mesons, respectively.  
For completeness, and to illustrate the effects of evanescent operators, we will present in this 
section complete one-loop results for all jet functions. 

The $B$-type currents in SCET$_{\rm I}$ may be written,
\begin{equation}\label{eq:JGamma}
  J_{\Gamma}(s,r) = \bar{\X}_{hc}(s\nb)\,\A_{hc\perp\mu}(r\nb)\,\Gamma_{\perp}^\mu\,h(0) \,,
\end{equation}
where the Dirac structure $\Gamma_\perp^\mu$ depends on the specific current 
under consideration; e.g. $\Gamma_\perp^\mu= \gamma_\perp^\mu$ for the scalar current.  
For the coefficient function of the operator $J_{\Gamma}(s,r)$, we write
\be\label{eq:CGamma}
  C_{\Gamma}(u) = \int\!dr\,ds\,e^{i(u s+\bar{u}r)\nb\cdot P}\, \tilde{C}_{\Gamma}(s,r) \,.
\end{equation}
The four-quark SCET$_{\rm II}$ operators depicted in Figure~\ref{fig:matching} take the form, 
\begin{eqnarray}\label{eq:fourquark}
  O^{kl,mn}_{\Gamma,i}(s,t) &=& \bar{\X}^k_c(s\nb) \Gamma_{\bar{\xi} h, i} \H^l(0)\, \bar{\Q}_s^m(tn) \Gamma_{\bar{q}\xi, i} \X_c^n(0) \,.
\end{eqnarray}
The superscripts $k,l,m, n$ are color indices, which will be suppressed in the following.  
Since we are interested in matrix elements for color-singlet initial and final state hadrons,
our primary concern is the color-singlet contribution, 
proportional to the contraction of (\ref{eq:fourquark}) with $\delta^{kn}\delta^{ml}$.   
Here $\X_c=W_c^\dagger \xi_c$ is the restriction of $\X_{hc}$ to collinear momentum modes. 
$\H=S^\dagger h$ and $\Q_s = S^\dagger q_s$ are soft fields multiplied by a soft Wilson line $S$ 
in the $n$-direction. 
The precise definitions of these fields and the SCET$_{\rm II}$ effective Lagrangian
is given in 
\cite{Becher:2003qh}. 

Unlike in the first matching step, from QCD onto SCET$_{\rm I}$, the matching of SCET$_{\rm I}$ 
onto SCET$_{\rm II}$ generates an infinite tower of SCET$_{\rm II}$ operators.  
From the Feynman rules of SCET$_{\rm I}$ it
follows that $\Gamma_{\bar{q}\xi}$ must take the form
\begin{equation}
  \Gamma_{\bar{q}\xi} = \gamma_{\perp\mu},\,
  \gamma_{\perp\mu_1\mu_2\mu_3},\,
  \gamma_{\perp\mu_1\mu_2\mu_3\mu_4\mu_5},\, \dots \,.
\end{equation} 
We denote totally antisymmetrized combinations of Dirac
matrices by $\gamma_{\perp}^{\mu_1\dots\mu_n}= \gamma_{\perp}^{[\mu_1}\dots
\gamma_{\perp}^{\mu_n]}$.
A complete basis is obtained by considering the most general
$\Gamma_{\bar{\xi}h}$ with the appropriate number of Lorentz indices,
constructed from $\Gamma_\perp^\mu$ and $\gamma_\perp^\mu$ (recall that
we choose the heavy-quark velocity such that $v_\perp^\mu=0$). 
At one-loop order, three structures are possible:
\begin{eqnarray}\label{eq:J1J2J3}
  \Gamma_{\bar{\xi} h, i}\otimes \Gamma_{\bar{q}\xi, i} &=& \left\{ 
  \begin{array}{rcll}
    \Gamma_\perp^\mu &\otimes& \gamma_{\perp\mu} \,, \quad &i=1 \,, \\
    \gamma_\perp^\mu \gamma_{\perp\rho} \Gamma_\perp^\rho &\otimes& \gamma_{\perp\mu} \,, \quad &i=2 \,, \\
    \gamma_\perp^{\mu\nu} \Gamma_\perp^\rho &\otimes& \gamma_{\perp\mu\nu\rho} \,, \quad &i=3 \,.
\end{array} 
\right.  
\end{eqnarray} 
Operators
for which $\Gamma_{\bar{q}\xi}$ involves antisymmetric products of five
and more Dirac matrices appear only at two and more loops. 
In four dimensions, antisymmetric products of more than 
two perpendicular Dirac matrices vanish.  
Operators such as $O_{\Gamma,3}(s,t)$ containing these Dirac
structures appear at one-loop and beyond in dimensional 
regularization, and are called evanescent~%
\cite{Buras:1989xd,Dugan:1990df,Herrlich:1994kh}.  
The choice of evanescent operators is not unique, and affects the coefficients of the
physical operators.  
We will discuss this issue, and the related subject of 
renormalization in the presence of evanescent operators in Section~\ref{sec:evanescent}. 
In this section we present results for the bare matching coefficients. 

We define the Fourier transform of the coefficient of the SCET$_{\rm II}$
operators $O_{\Gamma,i}(s,t)$ as
\begin{eqnarray}\label{eq:fourierD}
  D_{\Gamma,i}(\omega, u) = \int\! ds\,dt\, e^{-i\omega n\cdot v\, t} e^{iu s \nb\cdot P}\,
    \tilde{D}_{\Gamma,i}(s,t)\,,
\end{eqnarray}
where $\tilde{D}_{\Gamma,i}(s,t)$ is the position-space coefficient. 
The jet functions relating 
bare coefficients are defined via  
\begin{eqnarray}\label{eq:jetdef}
  D^{\rm bare}_{\Gamma,i}(\omega,u) = {1 \over  2E \omega} \int_0^1\! dv\, 
    {\cal J}_{i}(u,v, g^2(2E \omega)^{-\epsilon})\, C^{\rm bare}_{\Gamma}(v) \,,
\end{eqnarray}
which we expand as 
\begin{eqnarray}\label{eq:generaljet}
  {\cal J}_{i}(u,v,g^2(2E\omega)^{-\epsilon}) &=& 
    {g^2\over 2E\,\bar{u}}\bigg( -\delta_{i 1} \delta(u-v) ( {\cal S}+ {\cal O} )  \nl 
  && \qquad 
    + { g^2  \over (4\pi)^{2-\epsilon}}\Gamma(1+\epsilon) (2E\omega)^{-\epsilon} J^{(1)}_{i}(u,v) + \dots  \bigg) \,.
\end{eqnarray}
Here $g$ is the bare coupling; the dependence of ${\cal J}_i$ on $g$, $E$ and $\omega$ is determined by 
dimensional analysis and invariance under rescaling of the light-cone vector $n$.  
The symbols ${\cal S}$ and ${\cal O}$ represent the color-singlet and color-octet configurations of 
four-quark operators.
Restoring the color indices which are to be contracted with the operators in (\ref{eq:fourquark}), 
we define:
\bea
  {\cal S}^{kl,mn} &=& {C_F\over N} \delta^{kn}\delta^{ml} \nl
  {\cal O}^{kl,mn} &=& -{1\over N} (T_A)^{kn} (T_A)^{ml} \,,
\eea
where for convenience we include the tree-level factors in the definitions of ${\cal S}$ and ${\cal O}$.  

The one-loop jet-functions for the three operators in
(\ref{eq:J1J2J3}) are obtained from the SCET$_{\rm I}$
diagrams shown in Figure~\ref{fig:jet}. 
Diagrams involving soft 
modes need not be considered, since these momentum
regions are reproduced exactly by the corresponding SCET$_{\rm II}$ contributions, and hence
cancel in the matching.   
We proceed by evaluating the loop diagrams for an arbitrary
coefficient function $C_{\Gamma}$ in (\ref{eq:CGamma}).  
To perform the integration over the $d$-dimensional loop momentum $k$, 
we first integrate over the
$n\cdot k$ component using the method of residues,
and then perform the $k_\perp$ integral for the $d-2$ perpendicular dimensions. 
The remaining integration over $\bar{n}\cdot k$ takes the
form of equation (\ref{eq:jetdef}), from which we read off the contributions to the
jet-functions. In this way, we find
\begin{eqnarray} \label{eq:J1}
  J^{(1)}_{1}(u,v) &=& 
    \left[ {\theta(u-v)\over u-v}(u-v)^{-\epsilon}+{\theta(v-u)\over v-u}(v-u)^{-\epsilon} \right]_+  \bigg[ \left(C_F-{C_A\over 2}\right){\cal S} + \left(C_F-C_A\right){\cal O}\bigg]{2\over \epsilon} \nl
  && + \theta(u-v) (u-v)^{-\epsilon}\bigg[ \left(C_F-{C_A\over 2}\right){\cal S} + \left(C_F-C_A\right){\cal O}\bigg] \nl
  && \qquad \times {2\over u\bar{v}} \left[ -{1\over\epsilon} +{v\bar{u}\over u-v}\ln{u\over v} - {1\over 2}(1+v) \right] \nl
  && + \theta(v-u)(v-u)^{-\epsilon}\Bigg\{ \bigg[ C_F({\cal S}+{\cal O}) - {C_A\over 2}\left( {\bar{v}\over \bar{u}}{\cal S} + \left[ 1+{\bar{v}\over \bar{u}}\right]{\cal O}\right)\bigg] \nl
  && \qquad \times {2\over u\bar{v}}\left[ {u\bar{v}\over v-u}\ln{\bar{u}\over\bar{v}} - {1\over 2}(1+u) \right] +{C_A\over 2}({\cal S}+{\cal O}){2\over\bar{u}}\,{1\over \epsilon} \Bigg\}  \nl
  && + \left[ v\bar{v}\right]^{-\epsilon} {1\over u\bar{v}}\Bigg\{ C_F({\cal S}+{\cal O})\left[ (1+u)v-\bar{v}\right] -{C_A\over 2}\left[ v(1+u){\cal O}-\bar{v}{\cal S}\right] \Bigg\} \nl
  && + \theta(1-u-v)\left[{v(1-u-v)\over \bar{u}}\right]^{-\epsilon}
    \left(C_F-{C_A\over 2}\right) ({\cal S}+{\cal O}) {2v\over \bar{u}\bar{v}}\left( -{1\over \epsilon} +{\bar{v}\over uv} -{1\over v} \right) \nl
  && + \delta(u-v) u^{-\epsilon}\left[ \left(C_F-{C_A\over 2}\right){\cal S} + (C_F-C_A){\cal O} \right] \left(-{2\over \epsilon^2} \right) \nl
  && + \delta(u-v) \bar{u}^{-\epsilon} \Bigg\{ -{C_A\over 2}{\cal S}\left({2\over \epsilon^2} \right) + ({\cal S}+{\cal O})\bigg[\left({4\over 3}T_F n_f +3C_F -{11\over 3} C_A\right)\left({1\over \epsilon}\right) \nl
  && \qquad  
    + {20\over 9}T_Fn_f+8C_F -{76\over 9}C_A - (C_F-C_A){\pi^2\over 3}\bigg] \Bigg\} \,,
\end{eqnarray}

\begin{eqnarray} \label{eq:J2}
  J^{(1)}_{2}(u,v) &=& 
    \theta(u-v)\bigg[{(u-v)v\over u}\bigg]^{-\epsilon}\left[ \left(C_F-{C_A\over 2}\right){\cal S} + (C_F-C_A){\cal O}\right] {\bar{u}\over u\bar{v}}\left( {1\over \epsilon} + 1\right) \nl
  && 
    +  \theta(v-u)\bigg[{(v-u)\bar{v}\over \bar{u}}\bigg]^{-\epsilon}\left[ 
    C_F({\cal S}+{\cal O}) - {C_A\over 2}\left( {\bar{v}\over\bar{u}}{\cal S} + \left(1+{\bar{v}\over \bar{u}}\right){\cal O}\right)\right] 
    {1\over u}\left( {1\over \epsilon}+ 1\right) \nl
  && +  \left[ v\bar{v}\right]^{-\epsilon}\bigg[ 
    C_F({\cal S}+{\cal O}) {v\bar{u}\over u}\left(- {1\over \epsilon} + {\bar{v}\over v\bar{u}} \right)  
    +{C_A\over 2}\left( -{1\over u}{\cal S} + {\cal O}{\bar{u}\over u}\left[ {1\over \epsilon} - {u\over \bar{u}}\right] \right) \bigg] \nl
  && + \theta(1-u-v)\bigg[ {(1-u-v)v\over \bar{u}}\bigg]^{-\epsilon}
    \left( C_F- {C_A\over 2}\right) \left( {\cal S} + {\cal O}\right) \nl
  && \qquad  \times {1\over u\bar{u}\bar{v}}
    \left[ \left(uv-[1-u-v]^2\right){1\over \epsilon} - (1-u-v)(2-u-v)\right]\,,
\end{eqnarray}

\begin{eqnarray} \label{eq:J3}
  J^{(1)}_{3}(u,v) &=& 
    \theta(u-v)\bigg[{(u-v)v\over u}\bigg]^{-\epsilon}\left[ \left(C_F-{C_A\over 2}\right){\cal S} + (C_F-C_A){\cal O}\right] {1\over 2u}\left( {1\over \epsilon}+1\right)  \nl
  && + \theta(v-u)\bigg[{(v-u)\bar{v}\over \bar{u}}\bigg]^{-\epsilon}\left[ 
    C_F({\cal S}+{\cal O}) - {C_A\over 2}\left( {\bar{v}\over \bar{u}}{\cal S} 
    + \left(1+{\bar{v}\over \bar{u}}\right){\cal O}\right)\right] 
    {\bar{u}\over 2u\bar{v}}\left({1\over\epsilon} + 1\right) \nl
  && + \left[ v\bar{v}\right]^{-\epsilon}\bigg[ 
    C_F({\cal S}+{\cal O}){1-uv\over 2u\bar{v}}\left(- {1\over\epsilon}-1\right) \nl
  && \qquad 
    -{C_A\over 2}\left( {1\over 2u}{\cal S} + \left[{1\over 2u} + {1-uv\over 2u\bar{v}}\right]{\cal O} \right)\left(-{1\over\epsilon}-1\right) \bigg] \,.
\end{eqnarray}
The plus distribution appearing in $J^{(1)}_1$ is defined for symmetric functions $g(u,v)$ 
to act on test functions $f(v)$ as
\be\label{eq:plusDistrib}
  \int dv\,[g(u,v)]_+\,f(v) = \int dv\,g(u,v)\,\big[ f(v) - f(u) \big] \,.
\ee

We now return to the particular cases of the scalar, vector and
tensor current operators and extract their jet-functions from the 
general expressions above. 
Through one-loop order, one physical and one
evanescent four-quark operator are sufficient for the scalar case:
\begin{align}\label{eq:scalarops}
  O_{S1} &= \gamma_\perp^\alpha  \otimes \gamma_{\perp\alpha}\,,  &
  O_{S2} &= \gamma_\perp^{\alpha\beta\gamma} \otimes \gamma_{\perp\alpha\beta\gamma} \,.
\end{align}
To the same order, there are four physical and four evanescent vector operators
\begin{align}\label{eq:vectorops}
  O_{V1}^\mu &= {n^\mu\over n\cdot v} O_{S1} \,, & 
  O_{V2}^\mu &= {v^\mu} O_{S1}\,, &
  O_{V3}^\mu &= \gamma_{\perp}^\mu\gamma_\perp^\alpha \otimes \gamma_{\perp\alpha}\,,  &
  O_{V4}^\mu &= \gamma_{\perp}^\alpha\gamma_{\perp}^\mu \otimes \gamma_{\perp\alpha}\,, \nl
  O_{V5}^\mu &={n^\mu\over n\cdot v} O_{S2}\,, &  
  O_{V6}^\mu &=v^\mu O_{S2}\,, & 
  O_{V7}^\mu &= \gamma_\perp^{\alpha\beta}\gamma_\perp^\mu\gamma_\perp^\gamma \otimes \gamma_{\perp\alpha\beta\gamma}\,, &
  O_{V8}^\mu &= \gamma_\perp^{\alpha\beta\gamma}\gamma_\perp^\mu \otimes \gamma_{\perp\alpha\beta\gamma} \,, 
\end{align}
and a grand total of fourteen tensor operators
\begin{align}\label{eq:tensorops}
  O_{T1}^{\mu\nu} &= {n^{[\mu}v^{\nu]}\over n\cdot v} O_{S1} \,, &
  O_{T2}^{\mu\nu} &= {1\over n\cdot v} n^{[\mu} O_{V3}^{\nu]} \,, &
  O_{T3}^{\mu\nu} &= v^{[\mu} O_{V3}^{\nu]} \,, & 
  O_{T4}^{\mu\nu} &= {1\over n\cdot v}n^{[\mu}O_{V4}^{\nu]} \,, \nl 
  O_{T5}^{\mu\nu} &= v^{[\mu}O_{V4}^{\nu]} \,, & 
  O_{T6}^{\mu\nu} &= \gamma_\perp^\alpha\gamma_\perp^{\mu\nu} \otimes \gamma_{\perp\alpha} \,, &
  O_{T7}^{\mu\nu} &= \gamma_\perp^{\alpha\mu\nu} \otimes \gamma_{\perp\alpha} \,, &
  O_{T8}^{\mu\nu} &=  {n^{[\mu}v^{\nu]}\over n\cdot v} O_{S2} \,, \nl
  O_{T9}^{\mu\nu} &=  {1\over n\cdot v} n^{[\mu} O_{V7}^{\nu]} \,, &
  O_{T10}^{\mu\nu} &= v^{[\mu} O_{V7}^{\nu]} \,, &
  O_{T11}^{\mu\nu} &= {1\over n\cdot v} n^{[\mu} O_{V8}^{\nu]} \,, &
  O_{T12}^{\mu\nu} &= v^{[\mu} O_{V8}^{\nu]} \,, \nl
  O_{T13}^{\mu\nu} &= \gamma_{\perp}^{\alpha\beta\gamma}\gamma_\perp^{\mu\nu} \otimes \gamma_{\perp\alpha\beta\gamma} \,, &
  O_{T14}^{\mu\nu} &= \gamma_{\perp}^{\alpha\beta}\gamma_{\perp}^{\gamma\mu\nu} \otimes \gamma_{\perp\alpha\beta\gamma} \,,\hspace{-10mm}
\end{align}
of which the last eight are evanescent; in particular, $O_{T7}$, 
though appearing already at tree-level, is evanescent.%
\footnote{
Although three independent structures are possible for strings of five $\gamma_\perp$ matrices
with the appropriate antisymmetrizations, only $O_{T13}$ and $O_{T14}$ arise
in the matching.
}   
  As is evident from the above
equations many of these operators are directly related to each other,
e.g. $O_{V1}$ to $O_{S1}$, etc., and the matching onto these operators involves the same jet-functions. 
To obtain the jet-function in a particular  case, we first
rewrite the SCET$_{\rm I}$ current in the general form (\ref{eq:JGamma}), and then evaluate 
the three corresponding operators of (\ref{eq:J1J2J3}).  
Matching to the basis of four-quark operators defined in (\ref{eq:scalarops}), (\ref{eq:vectorops}) and (\ref{eq:tensorops})
yields the jet-functions in terms of the functions ${\cal J}_i$ defined in (\ref{eq:jetdef}). 
The independent matching relations are obtained from those operators with no factors of $v^\mu$ or $n^\mu$, and
with zero, one or two perpendicular Lorentz indices:  
\begin{align}\nonumber
 \begin{array}{ccc}
   & & \begin{array}{c}\phantom{1+} J_{S}^{B'}\phantom{1+2}
 \end{array} \\
 \bm{\cal J}_S =&
  \begin{array}{c}
  O_{S1} \\
  O_{S2} \\
  \vdots 
  \end{array} &
 \left(
  \begin{array}{c}
  {\cal J}_1 + 2(1-\epsilon){\cal J}_2 \\
  {\cal J}_3 \\
  \vdots 
  \end{array} 
 \right) \,,
  \end{array} 
&&
 \begin{array}{ccc}
  & & \begin{array}{ccc}\phantom{1+} J_{V1}^{B'}\phantom{1+2} & \phantom{123}J_{V4}^{B'} 
 \end{array} \\
 \bm{\cal J}_V^{\perp} =&
  \begin{array}{c}
  O_{V3} \\
  O_{V4} \\
  O_{V7} \\
  O_{V8} \\
  \vdots 
  \end{array} &
 \left(
  \begin{array}{cc}
   0 & {\cal J}_1  \\
   {\cal J}_1 + 2(1-\epsilon){\cal J}_2 \,\, & 2\epsilon {\cal J}_2 \\
   0 & {\cal J}_3 \\
   {\cal J}_3 & 0 \\
   \vdots & \vdots 
  \end{array} 
 \right) \,,
 \end{array} 
\end{align}
\begin{equation}\label{eq:jets}
 \begin{array}{ccc}
  & & \begin{array}{cc}\phantom{1+} J_{T1}^{B'}\phantom{1+2} &\phantom{1+2} J_{T5}^{B'} 
 \end{array} \\
 \bm{\cal J}_T^{\perp\perp} =&
  \begin{array}{c}
  O_{T6} \\
  O_{T7} \\
  O_{T13} \\
  O_{T14} \\
  \vdots 
  \end{array} &
 \left(
  \begin{array}{cc}
  {\cal J}_1 + 2(1-\epsilon){\cal J}_2 \,\, & -2\epsilon {\cal J}_2 \\
  0 & {\cal J}_1  \\
  {\cal J}_3 & 0 \\
  0 & {\cal J}_3 \\
  \vdots & \vdots 
  \end{array} 
 \right) \,.
 \end{array} 
\end{equation}
Since they are related to the operators above simply by overall factors of $v^\mu$ and $n^\mu$,
the remaining matching relations onto $O_{V1,2,5,6}$ and onto $O_{T1,\dots,5,8,\dots,12}$ 
may be obtained from (\ref{eq:jets}), using (\ref{eq:scalarops}), (\ref{eq:vectorops}) and (\ref{eq:tensorops}),  
and the analogous relations for the SCET$_{\rm I}$ currents in (\ref{eq:vectorprime}) and (\ref{eq:tensorprime}).  

From (\ref{eq:jets}), we note that  
for a particular SCET$_{\rm I}$ current $J_{\Gamma}$ in (\ref{eq:JGamma}), 
the index contractions in the resulting SCET$_{\rm II}$ operators
$O_{\Gamma,2}$ and $O_{\Gamma,3}$ in (\ref{eq:J1J2J3}) lead
to $\epsilon$-dependent prefactors, which need to be expanded around $d=4$ 
before renormalization in order to arrive at $\overline{\rm MS}$ subtracted jet-functions.
Before performing this renormalization, we first consider carefully the renormalization scheme dependence
implicit in our above choice of evanescent operator basis.   

\section{Evanescent Operators in SCET\label{sec:evanescent}} 

We have seen in the previous section that in dimensional
regularization, an infinite tower of evanescent operators is generated
upon matching SCET$_{\rm I}$ onto SCET$_{\rm II}$.  These operators
can mix into the physical basis, and so their effects must be included
for a consistent analysis.  
To ensure that these operators do not introduce additional nonperturbative
parameters into the theory, it is important to show that a
renormalization scheme can be chosen such that their renormalized matrix elements
vanish.  Here we show that this is in fact the case.  
A related issue concerns the choice of evanescent operator basis. 
As we will discuss, the renormalized matrix elements of physical operators  
depend on this basis choice, so that choosing a particular basis of
evanescent operators induces a renormalization scheme dependence in the physical 
operators.      
For the four-quark SCET$_{\rm II}$ operators defining the
hard-scattering part of the form factors, 
we choose the basis such that 
the resulting renormalization scheme corresponds to the 
$\overline{\rm MS}$ scheme after Fierz transformation.  
In Section~\ref{subsec:evan} we present a general discussion of evanescent operators 
in SCET, 
while in Section~\ref{subsec:appliedevan} we apply the general 
formalism to the operators appearing in the form factor analysis.  
    
\subsection{Evanescent operators and the Sudakov problem\label{subsec:evan}}

Evanescent operators first appeared in the case of local
four-fermion interactions~%
\cite{Buras:1989xd}.  
At one-loop order, it is
easy to see that the Green functions of bare evanescent operators are
finite and local, and hence can be made to vanish by the addition of a
suitable counterterm.  The finiteness of the Green functions results
from the $1/\epsilon$ pole from the momentum integral of the
corresponding Feynman diagrams being cancelled by a factor $\epsilon$
in the Dirac algebra; this latter factor of $\epsilon$ always appears
when evanescent operators ``cross the threshold'' and mix into
physical operators, reflecting the fact that evanescents, and hence
any contractions involving them, vanish in $d=4$.  The
locality of these particular Green functions is a consequence of
the locality of the coefficient of the $1/\epsilon$ pole of one-loop
integrals. 
A renormalization scheme in which matrix elements of evanescent operators
vanish to all orders in perturbation theory may then be established by
induction~%
\cite{Dugan:1990df}. 
In the present situation, where Green functions of the (nonlocal)
SCET current operators contain $1/\epsilon^2$ double poles already at
one-loop order, one may ask if an analogous renormalization scheme for
evanescent operators exists.  We will show that this is indeed the
case, both for SCET$_{\rm I}$ and SCET$_{\rm II}$. In practical
applications where we will ultimately match onto
SCET$_{\rm II}$, we may simply use the conventional $\overline{\rm MS}$
scheme also for evanescent SCET$_{\rm I}$ operators and perform finite
subtractions only in the final low energy theory. 
This is the procedure adopted in Section~\ref{subsec:appliedevan}. 

Let us consider first the simpler
case of the evanescent SCET$_{\rm I}$ current operator, $J_{T5}^{B'}$,
denoted by $J_{T5}$ in the following.  
Under renormalization,
this operator mixes with $J_{T1}^{B'}$, denoted by
$J_{T1}$.%
\footnote{This mixing only begins at two-loop order.}  
We proceed to isolate the linear combination of $J_{T1}$ and $J_{T5}$ 
which corresponds to the physical renormalized operator.  
It will be convenient to define the following notation for these
operators, considered as bare quantities,
\bea
  \J_1(s,r) &=& J_{T5}^{B'\mu\nu}(s,r) = \bar\X(s\nb)\,\A_{\perp\alpha}(r\nb)\,\gamma_\perp^{\alpha\mu\nu}\,h(0) \,, \nl
  \J_2(s,r) &=& \bar\X(s\nb)\,\calAslash_{\perp}(r\nb)\, \gamma_{\perp\alpha}\gamma_\perp^{\alpha\mu\nu}\,h(0)  
    = -2\epsilon\, \bar\X(s\nb)\,\calAslash_\perp(r\nb)\,\gamma_\perp^{\mu\nu}\,h(0) 
    = -2\epsilon J_{T1}^{B'\mu\nu}(s,r) \,. \hspace{5mm}
\eea
From the Feynman rules of SCET$_{\rm I}$ and the projection properties of the spinor field
$\X$, it follows that the operators $\J_1$ and 
$\J_2$ close under renormalization~%
\cite{Hill:2004if}.  
Defining renormalized operators via $\J_i^{\rm ren}=\sum_j Z_{ij}\,\J_j^{\rm bare}$, 
 the anomalous dimension matrix takes the form
\be\label{eq:simplescheme}
  \bm{Z} = \left( \begin{array}{cc}
    Z_{11} & Z_{12} \\
    0 & Z_{22}
   \end{array} \right) \,, 
\ee
where, in the $\overline{\rm MS}$ scheme, it is understood that only pole terms in the dimensional
regulator $\epsilon=2-d/2$ are kept in $\bm{Z}-\bm{1}$.%
\footnote{ $Z_{22}$ is given by the
pole part of $Z_{11}+2(1-\epsilon)Z_{12}$~%
\cite{Hill:2004if}; this
fact will not be relevant to our analysis.  } 
If we now consider a
Green function with an insertion of the renormalized operator,
$\J_1^{\rm ren}$, the result will be the sum of two terms, the first
proportional to the tree-level matrix element of $\J_1$, and the
second proportional to the tree-level matrix element of $\J_2$.  The
renormalization constants of $\bm{Z}$ ensure that the coefficients for
both terms are finite.  But $\J_1$ is evanescent, and $\J_2$ contains
an explicit factor of $\epsilon$, so their tree-level matrix elements
vanish at $d=4$; since the coefficients have been made finite by
renormalization, the final result also vanishes in $d=4$.  Returning to our
original operators, we define
\bea\label{eq:evscheme}
 \left( 
  \begin{array}{c}
  J_{T5}^{\rm ren} \\
  J_{T1}^{\rm ren} 
  \end{array}
 \right) &\equiv& 
 \left( 
  \begin{array}{c}
  \J_1^{\rm ren} \\
  -{1\over 2\epsilon} \J_2^{\rm ren} 
  \end{array}
 \right) =
 \left( \begin{array}{cc}
    Z_{11} & -2\epsilon Z_{12} \\
    0 & Z_{22}
   \end{array} \right)
 \left(
  \begin{array}{c}
  J_{T5}^{\rm bare} \\
  J_{T1}^{\rm bare} 
  \end{array}
 \right) \,.  
\eea 
$J_{T1}^{\rm ren}$ and $J_{T5}^{\rm ren}$ are equal to their bare
counterparts at tree level, and have finite Green functions to all
orders in the loop expansion, so that (\ref{eq:evscheme}) defines a valid
renormalization scheme.  Clearly $J_{T5}^{\rm ren}$, being equal to
$\J_1^{\rm ren}$, has vanishing matrix elements in this scheme. 
We may note that the renormalization of the new basis 
involves a finite subtraction from the term $\epsilon Z_{12}$, 
even if the original matrix $\bm{Z}$ in (\ref{eq:simplescheme}) 
is defined in the $\overline{\rm MS}$ scheme.   

None of this argument depended on the order in the loop
expansion, or on the choice of renormalization scheme used to
renormalize $\J_1$ and $\J_2$.  Also, since the renormalized matrix
element of $J_{T5}$ vanishes regardless of renormalization scale, we
find that
\bea
  0 = {d\over d\ln\mu} \langle J_{T5} \rangle  &=& -\gamma_{T5,T5} \langle J_{T5} \rangle   - \gamma_{T5,T1} \langle J_{T1} \rangle 
    =  - \gamma_{T5,T1} \langle J_{T1} \rangle \,,
\eea
where the last equality follows from the vanishing of $\langle J_{T5} \rangle$. Since the matrix element of the 
physical operator is nonzero,
it follows that $\gamma_{T5,T1} =0$.  The evolution equation for the coefficient function of the physical operator is
therefore simply 
\bea
{d\over d\ln\mu} C_{T1} &=& \gamma_{T1,T1} C_{T1} \,. 
\eea
In particular, the evolution equation is independent of $C_{T5}$, 
and the coefficient of the renormalized evanescent operator is irrelevant in the calculation of 
physical processes. 

\begin{figure}
\begin{center}
\includegraphics[width=0.8\textwidth]{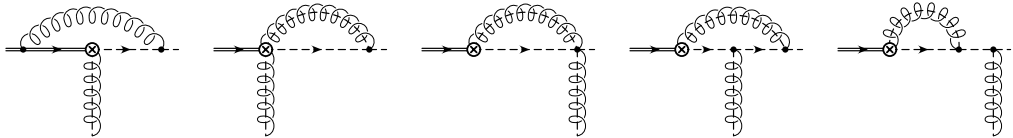}
\end{center}
\vspace{-0.5cm}
\caption{Abelian contribution to the anomalous dimension of the $B$-type $\mbox{SCET}_{\rm I}$ current operators. Full lines denote soft fields, dashed lines hard-collinear fields.\label{fig:abelian}}
\end{figure}

The simplicity of these arguments hides some interesting implications.
For instance, the vanishing of $\gamma_{T5,T1}$ implies that at
one-loop order $\epsilon Z_{12}$ is finite, or that $Z_{12}$ has at
most a single pole at $\epsilon=0$.  The one-loop $1/\epsilon^2$
double poles must occur only in the diagonal elements of $\bm{Z}$,
which do not mix the physical and evanescent operators.  To illustrate
how this works, we recall that the anomalous dimensions of the
$B$-type SCET$_{\rm I}$ current operators may be extracted from the UV
divergences of 3-point Green functions.  For simplicity we consider
the abelian case, for which the 
relevant diagrams are shown in Figure~\ref{fig:abelian}.  The soft contribution is
represented by the first diagram, while the remaining diagrams give 
the hard-collinear contribution.  Regulating infrared divergences by
evaluating the Green function at an off-shell value of the
hard-collinear quark momentum $p$, we find
\bea\label{eq:d1d2}
  \mbox{soft} &\propto& \left( -p^2\over 2E\, \mu \right)^{-2\epsilon} \left( -{1\over \epsilon^2} + \cdots \right) \,, \nl
  \mbox{hard-collinear} &\propto& \left( -p^2\over \mu^2 \right)^{-\epsilon} \left( {2\over \epsilon^2} + \cdots \right) \,. 
\eea
By general renormalization arguments~%
\cite{Collins:1984xc}, the
coefficient of the $1/\epsilon$ pole in the one-loop 3-point QCD Green function, which the
effective theory is representing, is local and hence independent of
$p^2$.  In fact, in SCET$_{\rm I}$, the QCD result is reproduced by
including the hard contribution, accounted for by the Wilson
coefficient, in addition to the hard-collinear and soft contributions.
This hard contribution is independent of $p^2$, but may depend on the
scales $m_b$ and $E$, so that the SCET$_{\rm I}$ divergences may be
nonlocal over distances $\sim 1/E$. This is exactly the structure obtained after 
taking the sum of the two contributions in (\ref{eq:d1d2}):
\bea\label{eq:sumd1d2}
  \mbox{soft + hard-collinear} &\propto&  {1\over \epsilon^2} - {2\over \epsilon}\ln{2E \over \mu} + \cdots \,,
\eea
where the remaining divergences are local at the level of SCET$_{\rm I}$, 
i.e. independent of $p^2$.  Since the Dirac structure of the original 
current is unaffected by the soft gluon exchange, the $1/\epsilon^2$ term in (\ref{eq:sumd1d2}) 
contributes only to the diagonal terms in $\bm{Z}$.  

We may present similar arguments to show that the renormalized matrix
elements of the evanescent SCET$_{\rm II}$ operators of the form 
(\ref{eq:fourquark}) vanish in $d=4$.  
Since the heavy quark couples at leading power only through a
spin-independent $v\cdot A_s$ interaction, no new Dirac structures can be
induced by the soft sector.  
Similarly, the soft-collinear modes couple to collinear modes only through $n\cdot A_{sc}$, 
and to soft modes only through $\nb\cdot A_{sc}$.  
All evanescent operators therefore arise
from the collinear sector, and we will only write the collinear part of the four-quark operators in our discussion. 
For simplicity, we restrict attention to the color-singlet case,%
\footnote{
For the color-singlet case, the soft-collinear modes
decouple entirely~%
\cite{Becher:2003kh}, 
so that the collinear and soft sectors
factorize.}  
and study the operators,
\bea
  [{O}_1]_{ab} = \left[ \bar{\X}\, \bar{\Gamma}_\perp \right]_a \,\, \left[ \Gamma_{\perp} \X \right]_b  \,, 
\eea
with spinor indices $a,b$. Here $\bar{\Gamma}_\perp$, $\Gamma_\perp$, 
represent as-yet unspecified Dirac structures ($\bar{\Gamma}_\perp$ is not related to
$\Gamma_\perp$).   
It follows from the Feynman
rules of SCET$_{\rm II}$ that ${O}_1$ mixes with the infinite tower of
operators,
\bea\label{eq:firstbasis}
  [{O}_n]_{ab} = \left[ \bar{\X}\, \gamma_\perp^{\nu_1}\cdots \gamma_\perp^{\nu_{2(n-1)}} \bar{\Gamma}_\perp \right]_a \,\, \left[ \Gamma_{\perp}\gamma_{\perp \nu_{2(n-1)}}\cdots \gamma_{\perp \nu_1} \X \right]_b  \,,
\eea
with ${O}_n$ appearing at $n$-th order in perturbation theory.  
Here the indices $\nu_i$ are not antisymmetrized.  
By general renormalization arguments~%
\cite{Collins:1984xc},
renormalization constants $Z_{nm}$ may be chosen so that with ${O}_n^{\rm ren} = \sum_m Z_{nm} {O}_m^{\rm bare}$, 
Green functions of the renormalized operators, ${O}_{n}^{\rm ren}$, are finite.  

In particular, in the representation of the scalar QCD current, we may consider the first evanescent operator, $O_{S2}$ 
in (\ref{eq:scalarops}),
for which we take
$\bar{\Gamma}_\perp\otimes \Gamma_\perp = \gamma_{\perp}^{\mu\rho\sigma}\otimes \gamma_{\perp\mu\rho\sigma}$.  
If we also introduce the operator
appearing at tree-level, 
\bea
  [{O}_0]_{ab}= \big[ \bar{\X}\, \gamma_\perp^\mu \big]_a \,\, \big[ \gamma_{\perp\mu} \X \big]_b  \,, 
\eea
then the renormalization matrix takes the form,%
\footnote{In fact, since at higher loop order the number of factors of $\gamma_\perp^{\nu_i}$ may not decrease, 
the matrix of renormalization constants is block triangular, with $Z_{nm}=0$ for $n<m$.  
We do not make use of this fact in the present discussion. 
}
\bea\label{eq:4quarkren}
 \left( 
  \begin{array}{c}
  {O}_0^{\rm ren} \\
  {O}_1^{\rm ren} \\
  {O}_2^{\rm ren} \\
  \vdots
  \end{array}
 \right)
 &=& 
 \left(
  \begin{array}{cccc}
  {Z}_{00} & {Z}_{01} & {Z}_{02} & \cdots \\
  0 & {Z}_{11} & {Z}_{12} & \cdots \\
  0 & {Z}_{21} & {Z}_{22} & \cdots \\
  \vdots & \vdots & \vdots & \ddots 
  \end{array}
 \right)
 \left( 
  \begin{array}{c}
  {O}_0^{\rm bare} \\
  {O}_1^{\rm bare} \\
  {O}_2^{\rm bare} \\
  \vdots
  \end{array}
 \right) \,. 
\eea
The reasoning now proceeds as in the previous case.  
The operators ${O}_n$ for $n\ge 1$ close under renormalization, 
and after renormalization, a Green function involving the insertion of one of these
operators is a linear combination of finite coefficients times tree-level matrix elements. 
Since the tree-level matrix elements vanish in $d=4$, the renormalized Green functions 
of the evanescent operators ${O}_n^{\rm ren}$, $n\ge 1$,  also vanish.  Again, since this
is true for any renormalization point, the elements of the anomalous dimension matrix 
describing the mixing of the physical operator into evanescents 
vanish.  Although we have specialized here to the case of the scalar current, the arguments may 
be easily extended to the general case. 

The basis of evanescent operators ${O}_n$ is of course not unique.  
We could for instance have chosen the antisymmetrized basis, 
\begin{align}\label{eq:antisymmbasis}
  [\hat{O}_n]_{ab} &= \big[ \bar{\X}\,   \gamma_\perp^{[ \nu_1}\cdots \gamma_\perp^{\nu_{2n+1}]} \big]_a \,\, 
    \big[ \gamma_{\perp [\nu_1}\cdots \gamma_{\perp\nu_{2n+1}]} \X \big]_b  \,,
\end{align}
or any other basis which is related to the original basis of operators by 
\begin{align}\label{eq:newbasis}
  {O}_0&=\hat{O}_0 &&\text{ and } &{O}_n &= \sum_{m=1}^n A_{nm} \hat{O}_m + \epsilon B_n \hat{O}_0 \;\text{ for }\; n\geq 1 \,,
\end{align}
with coefficients $A_{nm}$, $B_n$, finite at $d=4$.  The same 
renormalized operators may be expressed in terms of the 
new basis of bare operators as 
\bea
 \left( 
  \begin{array}{c}
  {O}_0^{\rm ren} \\
  {O}_1^{\rm ren} \\
  {O}_2^{\rm ren} \\
  \vdots
  \end{array}
 \right)
 &=& 
 \left(
  \begin{array}{cccc}
  {Z}_{00} + \epsilon \sum_{m=1}^\infty {Z}_{0m} B_m \,\, & \sum_{m=1}^\infty {Z}_{0m}A_{m1} \,\, 
  & \sum_{m=2}^\infty {Z}_{0m}A_{m2} & \cdots \\
  \epsilon \sum_{m=1}^\infty {Z}_{1m}B_m \,\,  & \sum_{m=1}^\infty {Z}_{1m}A_{m1} \,\,
  & \sum_{m=2}^\infty {Z}_{1m}A_{m2} & \cdots \\
  \epsilon \sum_{m=1}^\infty {Z}_{2m}B_m \,\,  & \sum_{m=1}^\infty {Z}_{2m}A_{m1} \,\,
  & \sum_{m=2}^\infty {Z}_{2m}A_{m2} & \cdots \\
  \vdots & \vdots & \vdots & \ddots 
  \end{array}
 \right) 
 \left( 
  \begin{array}{c}
  \hat{O}_0^{\rm bare} \\
  \hat{O}_1^{\rm bare} \\
  \hat{O}_2^{\rm bare} \\
  \vdots
  \end{array}
 \right) \!\!.  
\eea 
If we identify the resulting matrix with the renormalization constant
matrix $\hat{\bf Z}$ for the new basis, then we find for $n\ge 1$ 
the typical
counterterms, $\hat{Z}_{n0}= \epsilon \sum_{m=1}^\infty Z_{nm}B_m$, 
involving finite constants which must be included in order
to ensure the vanishing of the renormalized evanescent operators.  We
also find, however, a finite contribution to $\hat{Z}_{00}=Z_{00} +
\epsilon \sum_{m=1}^\infty Z_{0m} B_m$.  If, as is customary,
we impose the $\overline{\rm MS}$ scheme for the physical operators and
allow finite subtractions only for evanescent operators, in the terms
$\hat{Z}_{n0}$ for $n\ge 1$, then the finite contribution to ${\hat
Z}_{00}$ must be absorbed into the Wilson coefficient function of
$O_0^{\rm ren}$.  Since this finite contribution depends on the choice
of basis in (\ref{eq:newbasis}), a scheme dependence is induced in the
coefficient functions of the operators; this scheme
dependence cancels against the corresponding scheme dependence in the
operator matrix elements.

\subsection{Scheme dependence of the jet functions \label{subsec:appliedevan} }

From the preceding discussion, we find that using the $\overline{\rm
MS}$ scheme for the physical operators, a change of
evanescent operator basis corresponds to a change of renormalization
scheme.  In this section we work in the
opposite direction, and ask if it is possible to find a basis of
evanescent operators which yields a given renormalization scheme for
the physical operators.  In particular, when we take the matrix
elements of physical operators describing form factors, we would like
the LCDAs appearing in (\ref{eq:soft_plus_hard}) to be defined in the
$\overline{\rm MS}$ scheme.

We start by defining a Fierz rearranged operator $O'_i$ for each physical operator $O_i$  
in such a way that the bare operators coincide in four dimensions: 
\begin{align}\label{eq:FFierz}
  O_{S1}&=\bar{\X} \gamma_\perp^\alpha \H \,\, \bar{\Q}
    \gamma_{\perp\alpha} \X &
    &\leftrightarrow & 
  O'_{S1}&=2\,\bar{\X}_L {\sla{\nb}\over 2} \X \,\, \bar{\Q}_L
    {\sla{n}\over 2} \H+
    2\,\bar{\X}_R\, {\sla{\nb}\over 2}\, \X \,\, \bar{\Q}_R {\sla{n}\over
    2} \H \,, \nl 
  O_{V3}&=\bar{\X}
    \gamma_\perp^\mu\gamma_\perp^\alpha \H \,\, \bar{\Q}
    \gamma_{\perp\alpha} \X &
    &\leftrightarrow & 
  O'_{V3} &=2\bar{\X}_L \gamma_\perp^\mu {\sla{\nb}\over 2} \X \,\, \bar{\Q}_R {\sla{n}\over 2} \H + (L\leftrightarrow R)\,, \nl
  O_{V4} &=\bar{\X} \gamma_\perp^\alpha\gamma_\perp^\mu \H  \,\, \bar{\Q} \gamma_{\perp\alpha} \X &
    &\leftrightarrow &
  O'_{V4}&=2\bar{\X}_L  {\sla{\nb}\over 2}\X \,\, \bar{\Q}_L {\sla{n}\over 2} \gamma_\perp^\mu  \H + (L\leftrightarrow R) \,,\nl
  O_{T6} &=\bar{\X} \gamma_\perp^\alpha\gamma_\perp^{\mu\nu} \H  \,\, \bar{\Q} \gamma_{\perp\alpha} \X &
    &\leftrightarrow &
  O'_{T6}&=2\bar{\X}_L  {\sla{\nb}\over 2}\X \,\, \bar{\Q}_L {\sla{n}\over 2} \gamma_\perp^{\mu\nu}  \H + (L\leftrightarrow R)  \,,\nl
  O_{T7} &=\bar{\X} \gamma_\perp^{\alpha\mu\nu} \H  \,\, \bar{\Q} \gamma_{\perp\alpha} \X  &
    &\leftrightarrow &
  O'_{T7}&= - 2\bar{\X}_L \gamma_{\perp}^{[\mu} {\sla{\nb}\over 2}\X \,\, \bar{\Q}_R {\sla{n}\over 2} \gamma_\perp^{\nu]} \H + (L\leftrightarrow R) 
\,.
\end{align}
The remaining physical vector and tensor operators are obtained by
multiplying the above operators by factors of $v^\mu$ and $n^\mu$. In
fact, since $O_{V4}$ and $O_{T6}$ are obtained from $O_{S1}$ by
insertion of Dirac structures next to the heavy quark field, these
operators don't require an independent analysis.  Similarly, from the
identity
$-2\gamma_{\perp}^{\alpha\mu\nu}=\gamma_{\perp}^{\mu}\gamma_\perp^\alpha\gamma_\perp^\nu
-\gamma_{\perp}^{\nu}\gamma_\perp^\alpha\gamma_\perp^\mu $, we may
relate $O_{T7}$ to $O_{V3}$.
It is convenient to treat $O_{T7}$ in the same manner as the other operators in (\ref{eq:FFierz}),
even though it is evanescent.  We will find that the 
resulting renormalization scheme automatically ensures the vanishing of the renormalized $O_{T7}$.

\begin{figure}
\begin{center}
\begin{tabular}{ccc}
\raisebox{0.0\textwidth}{\includegraphics[width=0.13\textwidth]{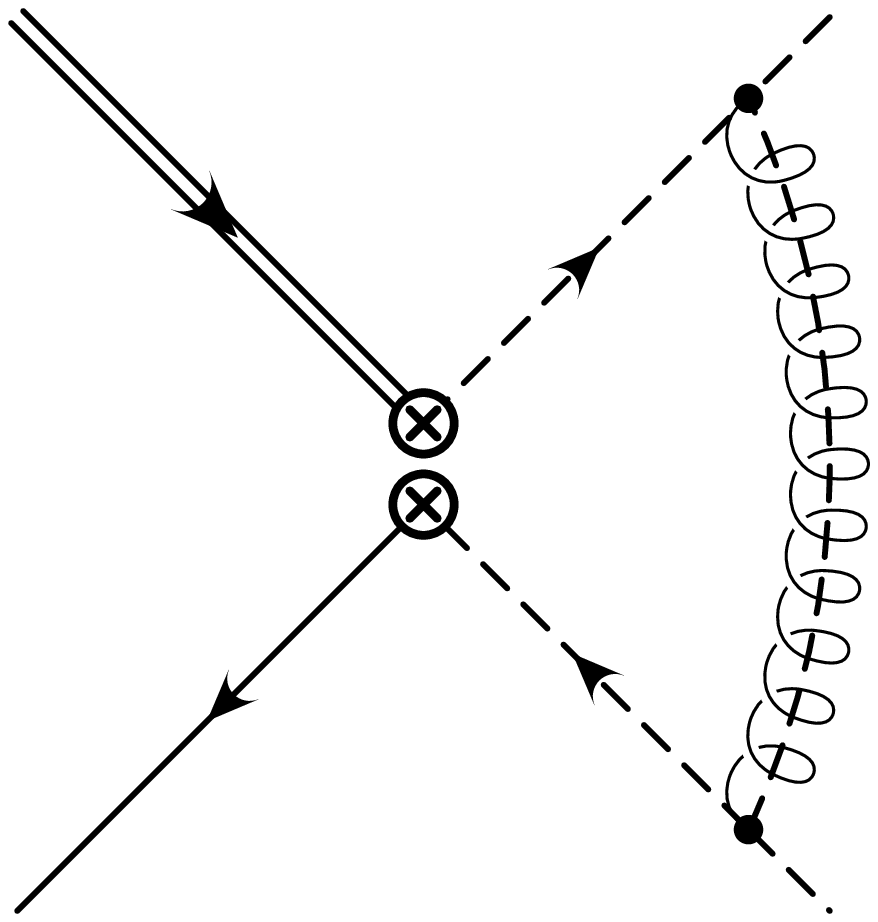}}
\phantom{aa}& \raisebox{0.07\textwidth}{\Large $-$} &
\raisebox{0.14\textwidth}{\includegraphics[width=0.13\textwidth,angle=270]{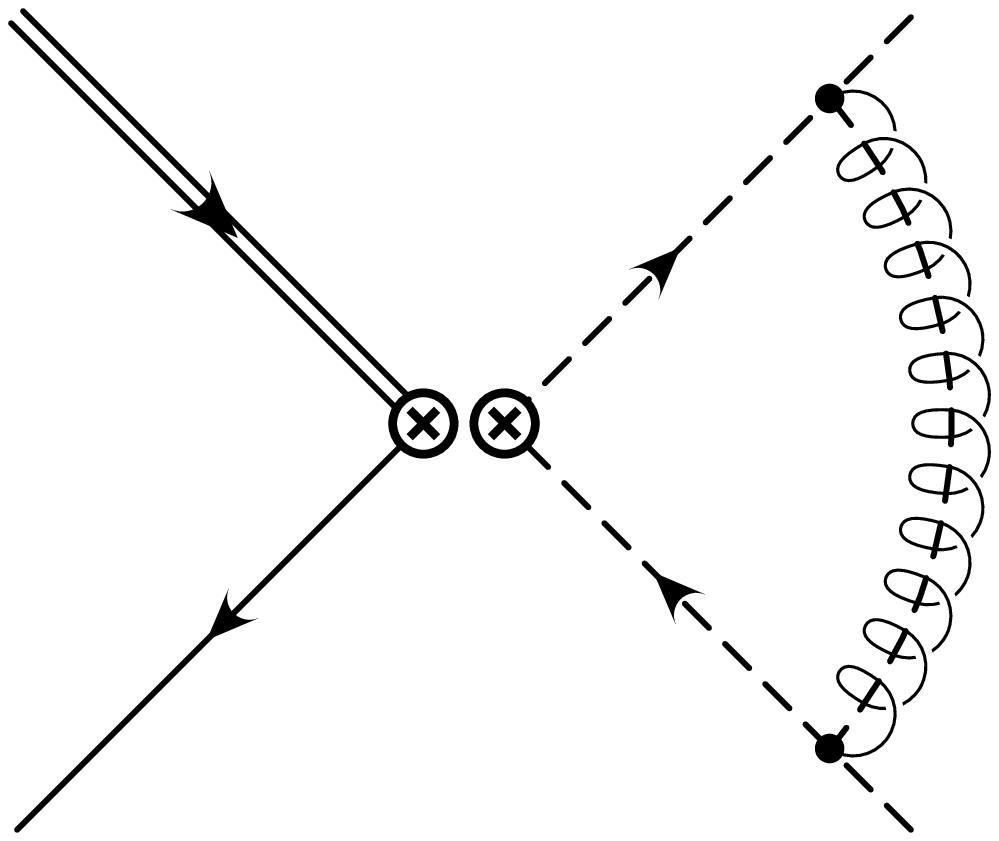}} 
\end{tabular}
\end{center}
\caption{One-loop graphs contributing to the difference between the matrix element 
of the original and the Fierz rearranged four-quark operators.} \label{fig:fierz}
\end{figure}
 
We define the renormalized 
physical operators $O_i$ and  $O_i'$ in the $\overline{\rm MS}$ scheme. The heavy-to-light meson matrix elements of the Fierz
rearranged operators $O_i'$ will
 then yield the product of the LCDAs of
the two mesons in the $\overline{\rm MS}$ scheme. Our task is to
define the evanescent operators in such a way that physical matrix
elements are the same for $O_i$ and $O'_i$.  For instance, if we
consider a four-point Green function involving $O_{S1}'$, then at
one-loop there are contributions proportional to the tree-level
structure, as well as a contribution proportional to
\bea\label{eq:primeloop}
  {1\over \epsilon} \times \gamma_\perp^\alpha\gamma_\perp^\beta  {\sla{\nb}\over 2} \gamma_{\perp\beta}\gamma_{\perp\alpha} 
    &=&  {1\over \epsilon} \times (4-8\epsilon + 4\epsilon^2) {\sla{\nb}\over 2}  
    \to -8  {\sla{\nb}\over 2}   \,. 
\eea
This contribution arises from the second diagram in
Figure~\ref{fig:fierz}.  In the $\overline{\rm MS}$ scheme, only the
$1/\epsilon$ pole is subtracted to obtain the renormalized result.
Similarly, the matrix element of $O_{S1}$ depicted in the first
diagram in Figure~\ref{fig:fierz} contains the same contribution
proportional to the tree-level structure $\gamma_\perp^\alpha\otimes
\gamma_{\perp\alpha}$, and a contribution proportional to
\bea\label{eq:loop}
  {1\over \epsilon} \times 
    \gamma_\perp^\alpha\gamma_\perp^\beta\gamma_\perp^\gamma \otimes \gamma_{\perp\gamma}\gamma_{\perp\beta}\gamma_{\perp\alpha} 
  &=&  {1\over \epsilon} \times (4-6\epsilon)\gamma_\perp^\alpha\otimes\gamma_{\perp\alpha} 
    - {1\over \epsilon} \times \gamma_\perp^{\alpha\beta\gamma} \otimes \gamma_{\perp\alpha\beta\gamma} \nl
  &\to& -6 \gamma_\perp^\alpha\otimes\gamma_{\perp\alpha} \,. 
\eea
Using $\overline{\rm MS}$ subtraction in the original basis (\ref{eq:scalarops}) yields a renormalized result 
different from the corresponding result in (\ref{eq:primeloop}).  
In place of (\ref{eq:loop}), we should instead write  
\bea 
  &&{1\over \epsilon} \times 
    \gamma_\perp^\alpha\gamma_\perp^\beta\gamma_\perp^\gamma \otimes \gamma_{\perp\gamma}\gamma_{\perp\beta}\gamma_{\perp\alpha} \nl
  &&=   {1\over \epsilon} \times (4-8\epsilon+4\epsilon^2) \gamma_\perp^\alpha\otimes\gamma_{\perp\alpha}  
    -{1\over \epsilon} \times \big[  \gamma_\perp^{\alpha\beta\gamma} \otimes \gamma_{\perp\alpha\beta\gamma} 
    - 2\epsilon(1-2\epsilon) \gamma_\perp^\alpha \otimes \gamma_{\perp\alpha} 
    \big] \nl
  &&\to -8 \gamma_\perp^\alpha\otimes\gamma_{\perp\alpha}   \,. 
\eea
The resulting contributions to the renormalized matrix elements of $O_{S1}'$ and $O_{S1}$ 
will therefore agree if we choose in place of $O_{S2}$ in (\ref{eq:scalarops}), 
\bea
  \hat{O}_{S2} = O_{S2} -2\epsilon(1-2\epsilon) O_{S1} \,. 
\eea
Beyond one-loop, the other evanescent operators $\hat{O}_{S3}, \hat{O}_{S4}, \dots$ are likewise determined.  
Proceeding in the same way for the vector case, we find 
\bea
  \hat{O}_{V7} &=& O_{V7} +2\epsilon(1+2\epsilon)O_{V3}  - 4\epsilon O_{V4} \,,\nl
  \hat{O}_{V8} &=& O_{V8} -2\epsilon(1-2\epsilon)O_{V4} \,.
\eea
Dropping the $\order(\epsilon^2)$ terms which are irrelevant through one-loop order, 
the scalar and vector jet-functions in this new basis are (cf. (\ref{eq:jets}))
\begin{align}\label{eq:scprime}
 \begin{array}{ccc}
  & & \begin{array}{c}\phantom{1+} J_{S}\phantom{1+2}
 \end{array} \\
 \hat{\bm{\cal J}}_S =&
 \begin{array}{c}
  O_{S1} \\
  \hat{O}_{S2} \\
 \end{array} &
 \left(
 \begin{array}{c}
  {\cal J}_1 + 2(1-\epsilon){\cal J}_2+2\epsilon {\cal J}_3 \\
  {\cal J}_3 \\
 \end{array} 
 \right) \,,
 \end{array} 
\end{align}
\begin{align}\label{eq:vecprime}
 \begin{array}{ccc}
  & & \begin{array}{ccc}\phantom{1+} J_{V1}^{B'}\phantom{1+2} & \phantom{123}J^{B'}_{V4} 
 \end{array} \\
 \hat{\bm{\cal J}}_V^{\perp} =&
 \begin{array}{c}
  O_{V3} \\
  O_{V4} \\
  \hat{O}_{V7} \\
  \hat{O}_{V8} \\
 \end{array} &
 \left(
 \begin{array}{cc}
  0 & {\cal J}_1-2\epsilon {\cal J}_3  \\
  {\cal J}_1 + 2(1-\epsilon){\cal J}_2+2\epsilon {\cal J}_3 \,\, & 2\epsilon {\cal J}_2 + 4\epsilon {\cal J}_3 \\
  0 & {\cal J}_3 \\
  {\cal J}_3 &  0 \\
 \end{array} 
 \right) \,.
 \end{array} 
\end{align}
Although they will not be relevant to the form factor analysis, for completeness we give here also the
new basis of evanescent operators for the tensor case, 
\bea
 \hat{O}_{T13} &=& O_{T13} -2\epsilon(1-2\epsilon) O_{T6} \,, \nl
 \hat{O}_{T14} &=& O_{T14} +2\epsilon(1+2\epsilon) O_{T7} +4\epsilon O_{T6} \,, 
\eea
and the corresponding jet functions, 
\begin{equation} \label{eq:newTjets}
 \begin{array}{ccc}
  & & \begin{array}{cc}\phantom{1+} J_{T1}^{B'}\phantom{1+2} &\phantom{1+2} J_{T5}^{B'} 
 \end{array} \\
 \hat{\bm{\cal J}}_T^{\perp\perp} =&
 \begin{array}{c}
  O_{T6} \\
  O_{T7} \\
  \hat{O}_{T13} \\
  \hat{O}_{T14} \\
 \end{array} &
 \left(
 \begin{array}{cc}
  {\cal J}_1 + 2(1-\epsilon){\cal J}_2 + 2\epsilon {\cal J}_3 \,\, & -2\epsilon {\cal J}_2 -4\epsilon {\cal J}_3 \\
  0 & {\cal J}_1 -2\epsilon {\cal J}_3  \\
  {\cal J}_3 & 0 \\
  0 & {\cal J}_3 \\
 \end{array} 
 \right) \,.
 \end{array} 
\end{equation}
We may recall from (\ref{eq:FFierz}) that the operators $O_{T7}$ and $O_{T7}^\prime$ vanish at tree level in $d=4$. 
In fact, since it is multiplicatively renormalized, the renormalized $O_{T7}^\prime$ then vanishes to all orders. 
Using the new basis of evanescents, $\hat{O}_{T13}, \hat{O}_{T14}, \dots$, the same is true for $O_{T7}$, 
since the new basis was chosen precisely to ensure that renormalized matrix elements of $O_{T7}$ and $O_{T7}^\prime$ agree.
In this scheme, therefore, $O_{T6}$ is identified with the renormalized physical operator, and the renormalized 
$O_{T7}$ vanishes.    
It is interesting to note that the evanescent $J_{T5}^{B'}$ gives a finite contribution at one-loop order 
to the bare matching coefficient of the physical operator $O_{T6}$ in (\ref{eq:newTjets}). 
As discussed in Section~\ref{subsec:evan}, since we use the  $\overline{\rm MS}$ scheme for all 
operators in SCET$_{\rm I}$, the corresponding 
matching coefficient between renormalized operators will also be nonzero.

Let us illustrate the above arguments by an explicit evaluation of the
diagrams in Figure~\ref{fig:fierz} for the scalar case.  In terms of
tree-level matrix elements, the difference between the renormalized matrix
elements at one-loop order is
\begin{multline}\label{eq:diff}
  \langle O_{S1}^{\rm ren}(u) \rangle(u^\prime) -\langle {O}_{S1}^{\prime\,\rm ren}(u) \rangle(u^\prime)
    =\int\!dv\, \Delta(v,u^\prime) \Big[  (4-6\epsilon) \langle O_{S1}(u)\rangle(v) -\langle O_{S2}(u)\rangle(v) \\  
  -4(1-\epsilon)^2\langle {O}_{S1}^\prime(u)\rangle(v) \Big]
    + \int\!dv\, Z_{S1,S2}(u,v) \langle O_{S2}(v)\rangle(u^\prime) \,. 
\end{multline}
Here $u$ is the Fourier transform variable in (\ref{eq:fourierD})
and $u'$ is the momentum fraction of the outgoing collinear quark in the diagrams in Figure~\ref{fig:fierz}.  
At tree level, we have $\langle O_i(u)\rangle(u^\prime) \propto \delta(u-u^\prime)$.  
The function $\Delta$ is given by 
\begin{equation}
  \Delta(u,u^\prime)
    =\frac{\alpha_s C_F}{4\pi}\frac{1}{2\epsilon}\left[\frac{u}{u^\prime}\theta(u^\prime-u)+\frac{\bar{u}}{\bar{u}^\prime}\theta(u-u^\prime)\right]
    =\frac{\alpha_s}{4\pi}\frac{(-\bar{u})}{\bar{u}^\prime} J_3^{(1)}(u^\prime,u)+O(\epsilon^0)\,,
\end{equation} 
where in the last equality we compare to the explicit form of $J_3^{(1)}$ in (\ref{eq:J3}).  
Since the divergences have to cancel, we see that $\Delta=Z_{S1,S2}$, 
and in our original evanescent operator basis the finite difference in the renormalized matrix elements becomes
\begin{align}
  \langle O_{S1}^{\rm ren}(u) \rangle(u^\prime) -\langle {O'}_{S1}^{\rm ren}(u) \rangle(u^\prime)
    &= \int\! dv\, 2\epsilon \Delta(v,u^\prime) \langle O_{S1}(u) \rangle(v)  \,.
\end{align}
The one-loop Wilson coefficient functions of these operators must therefore fulfill the relation
\bea\label{eq:scalarExpl}
  {\cal J}_{S1}'(u,v)-{\cal J}_{S1}(u,v)=\int\!dy\, {\cal J}_{S1}^{\rm tree}(y,v)\, 2\epsilon\Delta(y,u) =
    2\epsilon {\cal J}_3(u,v) \,, 
\eea
which is in agreement with (\ref{eq:jets}) and (\ref{eq:scprime}), since ${\hat{\cal J}}_{S1}={\cal J}_{S1}'$.

Having determined the operator basis, we now renormalize the
jet-functions.  After $\overline{\rm MS}$ subtractions, the quantities
${\cal J}_1+2(1-\epsilon){\cal J}_2 + 2\epsilon {\cal J}_3$ and ${\cal
J}_1-2\epsilon{\cal J}_3$ yield ${\cal J}_\parallel$ and ${\cal
J}_\perp$ from 
\cite{Hill:2004if}, 
respectively. We recall that ${\cal
J}_\parallel$ is the jet function appearing in all form factors
describing $B$-decays into pseudoscalar mesons, or
longitudinally-polarized vector mesons, while ${\cal J}_\perp$ appears
for all perpendicularly-polarized vector mesons.  In
\cite{Hill:2004if} 
we introduced the following notation,
\bea\label{j12def}
   {\cal J}_{\parallel,\perp}(u,v,L,\mu)
   &=& {4\pi C_F\alpha_s(\mu)\over N}\,{1\over 2E\bar u} \left[
    - \delta(u-v) + \frac{\alpha_s(\mu)}{4\pi}\,
     j_{\parallel,\perp}(u,v,L)  + \order(\alpha_s^2) \right] \,, 
\eea
where $L=\ln(2E\omega/\mu^2)$, $j_\parallel= j_1 + j_2 + j_3 $ and $j_\perp= j_1 - j_3$.  
In terms of the one-loop expressions
(\ref{eq:J1}) and (\ref{eq:J2}), after $\overline{\rm MS}$ subtractions we have  
\bea
  J_1^{(1)} &\to& j_1 \nl
  2(1-\epsilon) J_2^{(1)} &\to& j_2 \nl
  2\epsilon J_3^{(1)} &\to& j_3 \,.
\eea
The quantity $2\epsilon {\cal J}_2$ also appears, and we find 
\bea
  2\epsilon J_2^{(1)}(u,v) &\to& 
    2\left( C_F- {C_A\over 2}\right)\Bigg[ {\bar{u}\over u\bar{v}}\theta(u-v) + {\bar{v}\over u\bar{u}}\theta(v-u) 
    \nl
  && \quad + { uv - (1-u-v)^2 \over u\bar{u}\bar{v}} \theta(1-u-v) \Bigg] 
    + 2C_F \Bigg[ {v-u\over u\bar{u}}\theta(v-u) - {v\bar{u}\over u} \Bigg] \,. 
\eea
However, this jet function occurs only in the matching onto operators irrelevant to the form
factor analysis.  Similarly, the quantity ${\cal J}_3$ appears in the matching onto evanescents.  

\section{Application to heavy-to-light form factors\label{sec:app}}

\begin{figure}
\label{figure:typeA}
\psfrag{a}{$C^A_{F_+}, C^A_{T_1}$}
\psfrag{b}{$C^A_{F_0}$}
\psfrag{c}{$C^A_{F_T}$}
\psfrag{d}{$C^A_{A_1}$}
\psfrag{x}[]{$x=\frac{2E}{m_b}$}
\psfrag{y}{$C^A_i$}
\begin{center}
\includegraphics[width=0.8\textwidth]{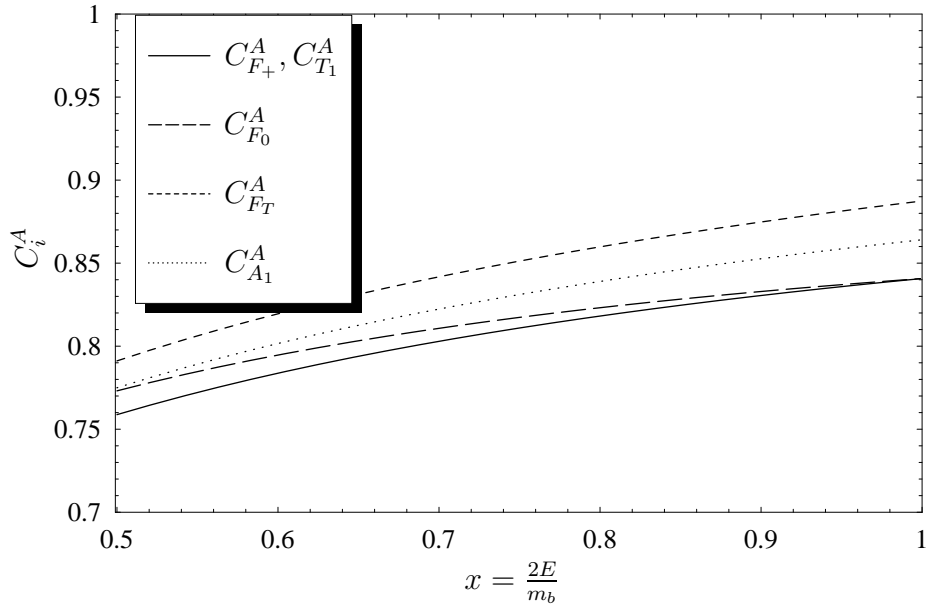}
\end{center}
\vspace{-0.5cm}
\caption{One-loop coefficients $C^A_i(E)$ of the soft-overlap terms in the
heavy-to-light form factors as a function of the recoil energy. The QCD as
well as the effective theory renormalization scale are set equal to $m_b$,
and we take $\alpha_s(m_b)=0.22$.   
All coefficients are equal to one at tree level.}
\end{figure}

In this section we study the numerical size of the matching
 corrections in the factorization formula for the form factors.  To
 begin, we review the situation for the $A$-type currents. This case
 is especially simple because no convolution over momentum fractions
 is involved, so that no assumptions on the shape of hadronic wave
 functions are necessary in order to make statements about the size of
 one-loop corrections relative to the tree-level values.  We consider
 the combinations of Wilson coefficients for the $A$-type currents
 which yield the $C_i$ in the first term of (\ref{eq:soft_plus_hard})
 for the form factors $F_+$, $F_0$, $F_T$, $V$ and $T_1$. To distinguish
 these from the $C^B_i$ in the hard-scattering terms, we will denote
 them by $C^A_i$ in the following. Form factor normalizations are
 chosen such that $C^A_{i}$ is equal to unity at tree level. The
 coefficients for the form factors of longitudinally-polarized vector
 mesons ($A_0$, $V-A_2$, $T_2-(2E/m_B)T_3$) are identical to the
 pseudoscalar case.  Also, for perpendicularly-polarized vector
 mesons, form factors $V$ and $T_2$ are related to $A_1$ and $T_1$
 respectively.  The same relations also hold for the hard-scattering
 terms.  Using the expressions from \cite{Bauer:2000yr,Beneke:2004rc},
 we find that the one-loop corrections are remarkably similar for all
 form factors. As illustrated in the plot in
 Figure~\ref{figure:typeA}, the corrections are negative, ranging in size from approximately 
$10-15\%$ at $E=m_b/2$ to $20-25\%$ at $E=m_b/4$. The symmetry breaking part of
 these corrections is very small, less than $5\%$ in the energy range
 considered.  This may indicate that the bulk of the one-loop correction is simply 
due to the choice of renormalize scheme or factorization scale.  
    
The analysis is more complicated for the $B$-type current operators.
Since the hard-scattering kernel $T_i(E,\omega,u)$ appearing in the
factorization formula (\ref{eq:soft_plus_hard}) depends on variables
$\omega$ and $u$, the numerical impact of the loop corrections can be
assessed only after convolution over these variables.  Knowledge of
the wavefunctions of the $B$-meson and light final state meson
appearing in the convolution is therefore necessary.  To begin, we
recall the form of the hard-scattering term~%
\cite{Hill:2004if}
\begin{multline}\label{eq:hardscatter}
  \Delta F_i = {m_B \over 2E}{f_B\over 4 K_F(\mu)} 
    \int_0^\infty {d\omega\over \omega}\phi_B(\omega,\mu) \int_0^1\!du\, f_M(\mu)\phi_M(u,\mu) \\
  \int_0^1 dv {\cal J}_\Gamma\left(u,v,\ln{2E\omega\over \mu^2},\mu\right) C^B_i(E,v,\mu)  \,,
\end{multline}
where ${\cal J}_\Gamma={\cal J}_\parallel$ for pseudoscalar and
longitudinally-polarized vector mesons and ${\cal J}_\Gamma={\cal J}_\perp$
for perpendicularly polarized vector mesons. The quantity
$K_F(\mu)=1+\frac{C_F \alpha_s}{4\pi}(3\ln\frac{m_b}{\mu}-2)$ relates the
QCD and the HQET $B$-meson decay constants. In order to avoid large
perturbative logarithms in the coefficients $C^B_i$, one would like to
set $\mu^2\sim m_b^2$, whereas the scale associated with the
jet-function is lower, $\mu\sim \sqrt{m_b\Lambda_h}$, where
$\Lambda_h$ is a typical hadronic scale. Since the form factor involves
different scales, we see that we cannot avoid large perturbative
logarithms in a fixed order calculation. These logarithms can be
resummed by solving the renormalization group equations for
the Wilson coefficients. This is discussed in detail in
\cite{Hill:2004if}, 
where we resummed all leading logarithms of
the different energy scales in the problem. The two effects we wish to
study here are the one-loop corrections to $C^B_i$, and to ${\cal
J}_\Gamma$. We study the two corrections in turn, evaluating each of them at
their natural scale $\mu$, but remind the reader that such a scale choice is
strictly only possible after renormalization group improvement.

\begin{figure}
\psfrag{a}{$C^B_{F_+}$}
\psfrag{b}{$C^B_{F_0}$}
\psfrag{c}{$C^B_{F_T}$}
\psfrag{d}{$C^B_{A_1}$}
\psfrag{e}{$C^B_{T_1}$}
\psfrag{x}[]{$x=\frac{2E}{m_b}$}
\psfrag{y}[B]{$C_i^{B(\rm eff)}- C_i^{B(\rm tree)}$}

\begin{center}
\includegraphics[width=0.8\textwidth]{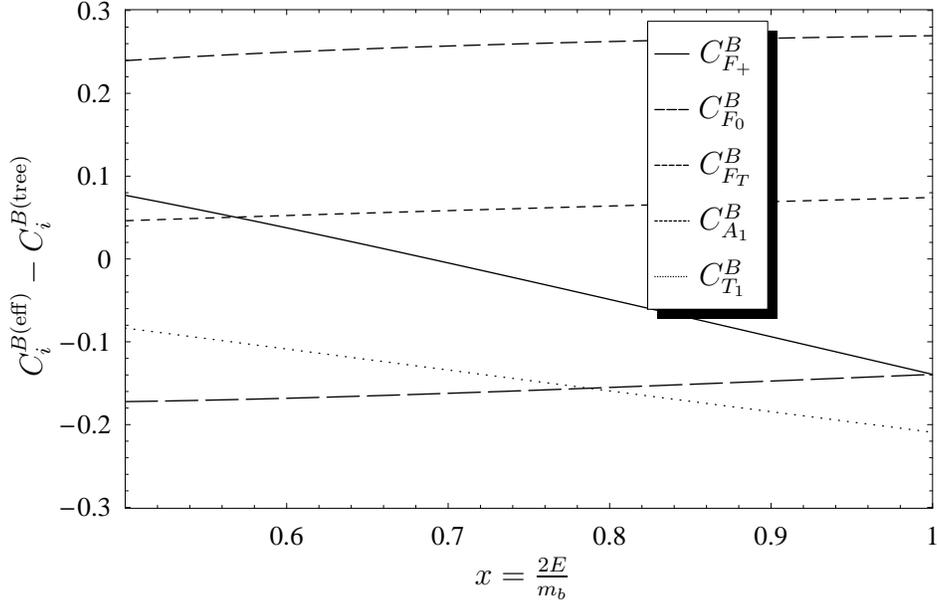}
\end{center}
\vspace{-0.5cm}
\caption{One-loop corrections to the form factors from the Wilson
coefficients $C^B_i(E,u)$. The figure shows the one-loop correction to
the quantity $C_i^{B(\rm eff)}(E)$ that arises after convolution of the
coefficient $C^B_i(E,u)$ with the leading order jet-function and the
asymptotic pion LCDA, see (\ref{eq:convB}).
The renormalization scale is set equal to $m_b$ in both QCD and the effective 
theory.\label{figure:typeB}}
\end{figure}

Let us start by considering the  hard-scale matching corrections to the $B$-type
Wilson coefficients.  The five independent combinations appearing in the form factors are
\begin{align}\label{eq:treeB} 
  C^B_{F_+} &={E\over m_b}C^{B'}_{V2} + C^{B'}_{V3} = 1-{4E\over m_b} + \dots\,, \nl
  C^B_{F_0} &=\left(1-{E\over m_b}\right)C^{B'}_{V2} + C^{B'}_{V3} = -1 + \dots\,, \nl
  C^B_{F_T} &= \frac12 C^{B'}_{T4} = 1 + \dots\,, \nl
  C^B_{A_1} &=C^{B'}_{V4} = 0 + \dots\,, \nl 
  C^B_{T_1} &= -\frac12 \left(1-{E\over m_b}\right)C^{B'}_{T6} -\frac12 C^{B'}_{T7} = -{2E\over m_b} + \dots \,.   
\end{align}
At tree level they take the values indicated; 
the pertinent one-loop expressions were given in Section~\ref{sec:scetI}. 
To evaluate the impact of these corrections on the form factors we work with the tree-level jet-function,
which is independent of $\omega$:  
\begin{equation}
  {\cal J}_\Gamma(u,v) =-\frac{4\pi\,C_F\,\alpha_s(\mu)}{N}\frac{1}{2E\bar u}\delta(u-v)+{\mathcal O}(\alpha_s^2) \,. 
\end{equation}
The size of the hard-matching corrections is then determined by convoluting the Wilson coefficient $C^B_i$ with the
light-meson wave function. Using for simplicity the asymptotic form
$\phi_M(u)=6u\bar{u}$ of the light meson LCDA, the relevant integral is
\begin{equation}\label{eq:convB}
  C_i^{B ({\rm eff})}(E,\mu) = \frac{2}{K_F(\mu)} \int_0^1{du} \,u\, C^B_i(E,u,\mu)\,. 
\end{equation}
These effective coefficients coincide with the $C^B_i$ at tree level,
since dependence on the momentum fraction $u$ appears only at one-loop order.  
In terms of $C_i^{B (\rm eff)}$ the correction to the form-factor is
\begin{equation}\label{eq:Fihard}
  \Delta F_i = - \left(m_B\over 2E\right)^2 \left[ {3\pi C_F\alpha_s(\mu)\over N} {f_B f_M(\mu)\over m_B \lambda_B(\mu)} \right]  C_i^{B(\rm eff)}(E,\mu)\,,
\end{equation}
where $\lambda_B(\mu)$ is defined as the first inverse moment of the $B$-meson LCDA,
\begin{equation}
  \lambda_B^{-1}(\mu) \equiv \int_0^\infty {d\omega\over \omega}
\phi_B(\omega,\mu) \,.  
\end{equation} 
The quantity inside square brackets in (\ref{eq:Fihard}) coincides, up to RG factors, with the
tree-level value of the function $H_M$ introduced in 
\cite{Hill:2004if}.  
In Figure~\ref{figure:typeB}, we plot the one-loop contributions to $C_i^{B (\rm eff)}(E,\mu)$. Choosing the scale $\mu=m_b$,
we find that the one-loop corrections are of order 20\% of the
tree-level values given in (\ref{eq:treeB}).

Finally, to isolate the one-loop corrections from the jet functions,
we use tree-level matching for the hard coefficients.  Since these
tree-level coefficients are independent of the momentum fraction $u$, in
this approximation the resulting convolution integrals in
(\ref{eq:hardscatter}) are universal to all form factors involving the
same final-state meson. For pseudoscalar and longitudinally-polarized
vector mesons, the result is proportional to~%
\cite{Hill:2004if}
\bea\label{eq:integrated_jet}
   &&\int_0^\infty\!{d\omega\over\omega}\,\phi_B(\omega,\mu)
    \int_0^1\!{du}\,\phi_M(u,\mu)
    \int_0^1\!dv\,{\cal J}_\parallel(u,v,\ln(2E\omega/\mu^2),\mu) \quad \nl
   &&\propto 1 + {\alpha_s(\mu)\over 4\pi}
    \left[ {4\over 3} \left\langle \ln^2{2E\omega\over\mu^2} \right\rangle
    + \left( - {19\over 3} + {\pi^2\over 9} \right)
    \left\langle \ln{2E\omega\over \mu^2} \right\rangle + 3.99 \right] \,,
\eea
where for simplicity we have again used the asymptotic form for the
light-meson LCDAs.  
We take $n_f=4$ as the number of light quark flavors. 
For perpendicularly-polarized vector mesons, where
${\cal J}_\parallel$ is replaced by ${\cal J}_\perp$, the coefficient
of the double logarithm is the same as in (\ref{eq:integrated_jet}),
the coefficient of the single logarithm is changed from
$(-19/3+\pi^2/9)$ to $(-6+\pi^2/9)$, and the constant term is changed from
$3.99$ to $0.73$.  Here the angle brackets denote averages over the
$B$-meson LCDA,
\bea\label{eq:wfav}
  \langle g(\omega) \rangle \equiv \lambda_B(\mu) \int_0^\infty{d\omega\over\omega} \phi_B(\omega,\mu)\, g(\omega) \,.
\eea
To investigate the size of the one-loop corrections, we may take
illustrative models for the $B$-meson LCDA. For example, in
\cite{Grozin:1996pq}, 
\be\label{eq:Bmodel}
   \phi_B(\omega,\mu_0) = {\omega\over\lambda_B^2}\,
   e^{-\omega/\lambda_B} \,, \qquad
   \lambda_B = {2\over 3}\,(m_B-m_b)\approx 0.32\,\mbox{GeV} \,, 
\ee 
where $\mu_0$ is a low hadronic scale at which the model is assumed valid.  
For the averages in (\ref{eq:integrated_jet}), we evaluate the wavefunction appearing in 
(\ref{eq:wfav}) at the scale $\mu_0$, and 
take the intermediate scale $\mu_i^2 =  m_b\Lambda_h$, with $\Lambda_h=0.5\,{\rm GeV}$, 
as the argument of the jet function.    
For maximal energy $E=m_b/2$ this model yields
\bea
  \left\langle \ln{2E\omega\over\mu_i^2} \right\rangle &=& \ln\left(2E\lambda_B\exp^{-\gamma_E}\over \mu_i^2\right) \sim -1.0 \,,\nl
  \left\langle \ln^2\left(2E\omega\over\mu_i^2\right) \right\rangle &=& \ln^2\left(2E\lambda_B\exp^{-\gamma_E}\over \mu_i^2\right) + {\pi^2\over 6} \sim 2.7 \,.
\eea
The resulting hard-scattering corrections in (\ref{eq:integrated_jet})
are approximately $+ 30\%$ and $ + 40\%$ for ${\cal J}_\parallel$ and ${\cal J}_\perp$ respectively.  
Neglecting the implicit model-dependence of these results, the uncertainty due to $m_b$ and hence of 
$\lambda_B$ is still significant; for instance, taking $\lambda_B = 0.46\,{\rm GeV}$~%
\cite{Braun:2003wx}
reduces the size of corrections to approximately $+ 20\%$ and $+30\%$.    
Using the model wavefunction from a more recent analysis in \cite{Braun:2003wx},
with the same value of the intermediate scale in the jet function, 
and at maximal energy $E=m_b/2$, gives 
\begin{align}
  \left\langle \ln{2E\omega\over\mu_i^2} \right\rangle &\sim -0.7 \pm 0.4\,,  &
  \left\langle \ln^2\left(2E\omega\over\mu_i^2\right) \right\rangle &\sim 1.0 \pm 0.6 \,,
\end{align}
where the wavefunction is evaluated at the scale $1\,{\rm GeV}$.   
In this model, the corrections for ${\cal J}_\parallel$ and 
${\cal J}_\perp$ are approximately $15\%$ and $25\%$ at $E=m_b/2$, and grow larger at small values of $E$, reaching  
$30\%$ and $40\%$ at $E=m_b/4$. 
In evaluating the averages in (\ref{eq:integrated_jet}) 
with wavefunctions evaluated at the low scale $\mu\sim 1\,{\rm GeV}$, and at the same
time taking $\mu_i = m_b\Lambda_h$ as the argument of the jet functions, 
we have neglected the RG factors relating the 
scales $\mu$ and $\mu_i$.  
From 
\cite{Hill:2004if}, 
we find the resulting correction to (\ref{eq:integrated_jet}) 
for this choice of scales to be approximately $-15\%$.

\section{Discussion and conclusions}\label{sec:conclusion}

We have presented in this paper the complete results for the one-loop
matching corrections to the hard-scattering coefficients in the
factorization formula (\ref{eq:soft_plus_hard}) which describes
heavy-to-light form factors at large recoil.  Our results
(\ref{eq:vectorB}) and (\ref{eq:tensorB}) for the hard-scale matching
coefficients $C^{B}_i$ in the vector and tensor case extend our
previous scalar current results reported in 
\cite{Hill:2004if}, 
and are in agreement with an independent calculation~%
\cite{Beneke:2004rc}.
Results for the intermediate-scale coefficients, ${\cal J}_\parallel$
and ${\cal J}_\perp$ in (\ref{j12def}), which are relevant to
$B$ decay form factors were reported in 
\cite{Hill:2004if}; 
in this paper we have presented general results, and details of the
calculation.

An interesting theoretical issue involves the evanescent operators
which arise upon matching QCD onto the effective theory.  We showed
that even in the presence of divergences of the Sudakov type,
a renormalization scheme exists for which renormalized evanescent
operators vanish.  The issue arises already in SCET$_{\rm I}$, for the
$B$-type operators, and we discussed this case first. This example 
is especially interesting   
since the mixing involves only two operators, in contrast
to the usual infinite tower of evanescent operators induced by
radiative corrections to 
four-fermion interactions.  The case of the nonlocal four-quark
operators of SCET$_{\rm II}$ was then treated along the same lines.  On
a practical level, use of such a renormalization scheme implies that
matching coefficients onto evanescent operators become irrelevant 
below the matching scale, 
since the renormalized operator
matrix elements vanish, and the physical operators do not mix into the
evanescents.  
We also discussed the choice of evanescent operator 
basis.  Different bases correspond to different renormalization schemes
for the physical operators.  We isolated a particular basis which corresponds
to the $\overline{\rm MS}$ scheme after Fierz transformation.  

We have presented an analysis of the phenomenological impact of
one-loop radiative corrections in Section~\ref{sec:app}.  Using the
known results for the $A$-type currents~%
\cite{Bauer:2000yr,Beneke:2004rc}, we find that the matching
corrections to the soft-overlap terms are remarkably universal to all
form factors.  The corrections are of order $15\%$, but differ in all
cases by less than $5\%$.  This effect 
seems not to have been noticed previously in the literature, and 
implies that any breaking of the
naive large-energy spin-symmetry relations by more than $5\%$ is due
to either the hard-scattering contributions, or to uncalculated power
corrections.  We have also investigated the numerical impact of
radiative corrections to the hard-scattering coefficients, finding the
hard-scale corrections to be $\lesssim 20\%$ for all form factors.
These corrections determine the size of violations to the universality
of the hard-scattering terms for different form factors which holds at
tree level for the hard-scale coefficient.  Radiative corrections to
the jet functions are less important from the point of view of
universality, since they contribute identically to all form factors
involving the same final state meson.  On the other hand, for the
purpose of calculating the hard-scattering contributions in terms of
the $B$ meson wavefunction, the jet function corrections are expected
to be more important than the hard-scale corrections, since the
coupling constant $\alpha_s(\mu)$ is larger at the intermediate scale
$\mu^2\sim\Lambda_{\rm QCD}m_b$.  For this reason it has been
speculated that the jet-functions might not have a convergent
perturbative expansion for physical values of $m_b$~%
\cite{Bauer:2004tj}.  We evaluated the perturbative corrections to the
jet-functions for two models of $\phi_B(\omega)$, finding that
one-loop corrections are of order $20-30\%$, with a large uncertainty
owing to the uncertainty in the model parameters.

We have focused here on the heavy-to-light meson form factors, but the
 same Wilson coefficients are also relevant for other
 processes. The jet-functions calculated here appear in other exclusive
 processes, such as $B\rightarrow K^*\gamma$ and the decays to two
 light mesons~%
 \cite{Bauer:2004tj}. 
The one-loop matching
 corrections to the SCET$_{\rm I}$ current operators might become
 relevant for the study of power corrections to inclusive decay
 distributions for processes like $B\rightarrow X_u \ell \nu$ and
 $B\rightarrow X_s \gamma$~%
\cite{Neubert:2004dd}.
 
\subsection*{Acknowledgments}

We thank M.~Beneke and D.~s.~Yang for alerting us to the scheme
dependence of the one-loop jet functions and M.~Neubert for useful
discussions. The work of T.B.\ and R.J.H.\ is supported by the
Department of Energy under Grant DE-AC02-76SF00515 and by the National
Science Foundation under Grant PHY99-07949.

\end{document}